\documentclass[12pt,preprint]{aastex}

\shorttitle{{\it Suzaku} Observations of HESS~J1731$-$347}
\shortauthors{Bamba et al.}

\begin{document}

\title{{\it Suzaku} Observations of the Non-thermal Supernova Remnant
HESS~J1731$-$347}

\author{
Aya Bamba\altaffilmark{1},
Gerd P{\" u}hlhofer\altaffilmark{2},
Fabio Acero\altaffilmark{3},
Dmitry Klochkov\altaffilmark{2},
Wenwu Tian\altaffilmark{4,5},
Ryo Yamazaki\altaffilmark{1},
Zhiyuan Li\altaffilmark{6},
Dieter Horns\altaffilmark{7},
Karl Kosack\altaffilmark{8},
Nukri Komin\altaffilmark{9}
}

\altaffiltext{1}{
Department of Physics and Mathematics, Aoyama Gakuin University
5-10-1 Fuchinobe Chuo-ku, Sagamihara,
Kanagawa 252-5258, Japan
}

\altaffiltext{2}{
Institut f{\" u}r Astronomie und Astrophysik,
Universit{\" a}t T{\" u}bingen,
Sand 1, D 72076 T{\" u}bingen, Germany
}

\altaffiltext{3}{
Laboratoire Univers et Particules de Montpellier, Universite
Montpellier 2, CNRS/IN2P3, CC 72, Place Eugene Bataillon,
F-34095 Montpellier Cedex 5, France}

\altaffiltext{4}{
National Astronomical Observatories, CAS, Beijing 100012, China
}

\altaffiltext{5}{
Department of Physics \& Astronomy, University of Calgary,
Calgary, Alberta T2N 1N4, Canada
}

\altaffiltext{6}{
Harvard-Smithsonian Center for Astrophysics,
60 Garden Street, Cambridge, MA 02138, USA
}

\altaffiltext{7}{
Universit{\" a}t Hamburg,
Institut f{\" u}r Experimentalphysik,
Luruper Chaussee 149, 22761 Hamburg, Germany 
}

\altaffiltext{8}{
CEA Saclay, DSM/IRFU, 91191 Gif-Sur-Yvette Cedex, France
}

\altaffiltext{9}{
Laboratoire d'Annecy-le-Vieux de Physique des Particules,
Universite' de Savoie, CNRS/IN2P3, F-74941 Annecy-le-Vieux,
France
}

\begin{abstract}
A detailed analysis of the nonthermal X-ray emission
from the North-Western and Southern parts of the supernova remnant (SNR)
HESS~J1731$-$347 with {\it Suzaku} is presented.
The shell portions covered by the observations emit hard and line-less X-rays.
The spectrum can be reproduced by a simple absorbed power-law model
with a photon index $\Gamma$ of 1.8--2.7
and an absorption column density $N_{\rm H}$ of
(1.0--2.1)$\times 10^{22}$~cm$^{-2}$.
These quantities
change significantly
from region to region;
the North-Western part of the SNR has the hardest and most absorbed
spectrum.
The Western part of the X-ray shell has a smaller curvature
than North-Western and Southern shell segments.
A comparison of the X-ray morphology to the Very High Energy (VHE) gamma-ray
and radio images was performed.
The efficiency of electron acceleration and emission mechanism
in each portion of the shell are discussed.
Thermal X-ray emission from the SNR was searched for but
could not be detected at a significant level.
\end{abstract}

\keywords{acceleration of particles
--- ISM: individual (\objectname{HESS~J1731$-$347})
--- X-rays: ISM}

\section{Introduction}

Supernova remnants (SNRs) are widely believed to be 
the main source of Galactic cosmic rays.
Observationally, several SNRs are found to have
X-ray synchrotron-emitting shells,
and are therefore widely known as electron accelerators
at least up to $\sim$TeV energies.
Many of them are identified as young SNRs
from the records of supernovae,
expansion velocity, and their bright thermal X-rays
(e.g. Cas~A, Tycho, Kepler, SN~1006; \cite{koyama1995,bamba2005a}).
However,
the nature of the prominent non-thermal X-ray-emitting SNRs
RX~J1713$-$3946 and Vela Jr. is still largely unclear
\citep{koyama1997,slane2001,bamba2005}.
The very low level of radio synchrotron emission,
the lack of thermal X-ray emission,
and poor expansion velocity data have yielded
only weakly constrained key parameters of the SNRs,
such as age and nature of the progenitor star.
Remarkably,
these SNRs are bright Very High Energy (VHE) gamma-ray emitters
compared with other young SNRs
\citep{aharonian2004,aharonian2005}.
Thus, they are strong particle accelerators,
and it is important to understand the nature of this small class of sources,
and especially the puzzling absence of thermal X-rays.

HESS~J1731$-$347 has been discovered
in the VHE Galactic plane survey performed with the H.E.S.S. telescopes
\citep{aharonian2008}.
\citet{tian2008} discovered a shell structure
in archival radio data spatially coincident with the VHE source,
and the source (also named G353.6$-$0.7) was hence considered
to be a newly discovered SNR
with associated VHE gamma-ray emission.
Subsequently, deep VHE follow-up observations were performed
\citep{abramowski2011} that established a clear VHE shell-like
structure in agreement with the radio shell.

Based upon a possible association of the SNR
with a nearby HII region,
\citet{tian2008} suggested that the object could have the same distance
($3.2\pm0.8$~kpc),
using a flat Galactic rotation curve model and
the most recent estimates of the parameters by
\citet{eisenhauer2005}.
A lower limit of the distance comes from the clear correlation
of the X-ray absorption column density with a molecular cloud structure
at $\sim$3~kpc as seen in CO emission \citep{abramowski2011},
suggesting that the cloud is in the foreground of or
interacting with the SNR.
This SNR has no GeV counterpart in the {\it Fermi} 2-year
Point Source Catalog \citep{fermi2011}.

{\it XMM-Newton} and {\it Suzaku} observations
towards the center and the East of this SNR show that
X-ray emission from the shell is dominated by
non-thermal emission
\citep{acero2009,tian2010}.
This suggests that
HESS~J1731$-$347 could be the third member of the SNR class
characterized by pure non-thermal X-ray emission
and comparable VHE gamma-ray luminosity.
A sensitive search for thermal X-ray emission
across the entire remnant is needed to confirm this hypothesis.

From spatially resolved X-ray synchrotron spectra,
changes of particle acceleration conditions can be studied.
Such variations have already been found 
in other SNRs such as RX~J1713$-$3946,
and were attributed to electron cooling over the distance from
the acceleration site \citep{tanaka2008}.

In this paper, we report on {\it Suzaku} observations
\citep{mitsuda2007}
of the western half of the SNR HESS~J1731$-$347.
The observation details are summarized in \S\ref{sec:obs}.
A first analysis of the {\it Suzaku} X-ray data is
presented in \S\ref{sec:results},
the results of which are discussed in \S\ref{sec:discussion}.

\section{Observations and Data Reduction}
\label{sec:obs}

{\it Suzaku} observed the center region of HESS~J1731$-$347
on 2007 Feb. 23--24,
and the North-Western and Southern regions on 2010 Feb. 17--19. 
The mapping consists of three pointings as listed in Table~\ref{tab:obslog}.
Data reduction and analysis were made
using HEADAS software version 6.9,
version 2.5.16.28 of the processed data,
and XSPEC version 12.6.0.

{\it Suzaku} has two active instruments,
four X-ray Imaging Spectrometers \citep[XIS0--XIS3;][]{koyama2007}
each at the focus of an X-Ray Telescope \citep[XRTs][]{serlemitsos2007}
and a separate Hard X-ray Detector \citep[HXD;][]{takahashi2007}.

Only three XISs could be used subsequently
due to a problem with XIS~2.
XIS1 is a back-illuminated (BI) CCD,
whereas the others are front-illuminated (FI).
The XIS instruments were operated
in normal full-frame clocking mode.
Spaced-row charge injection \citep{nakajima2008,uchiyama2009}
was used in the later two observations (Table~\ref{tab:obslog}).
We filtered out data acquired during passages through the
South Atlantic Anomaly (SAA),
with elevation angle to the Earth's dark limb below 5$^\circ$,
or with elevation angle to the bright limb below 25$^\circ$
in order to avoid contamination by emission from the bright limb.
The remaining exposure time for each observation is summarized in
Table~\ref{tab:obslog}.

The HXD PIN was operated in normal mode.
We filtered out data
obtained during passages through the
SAA,
with elevation angle to the Earth's limb below 5$^\circ$,
and cut off rigidity (COR) smaller than 8~GV.
The exposure time for each observation is again listed in
Table~\ref{tab:obslog}.
For the non-X-ray background (NXB) model,
we adopted the {\tt LCFIT} model by \citet{fukazawa2009}.
The cosmic X-ray background (CXB) flux is estimated
based on the {\it HEAO1} results \citep{boldt1987},
and treated as an additional background component.

\section{Results}
\label{sec:results}

\subsection{X-ray images}

Figure~\ref{fig:images} shows 
0.5--2.0~keV and 2.0--8.0~keV mosaic images
obtained with the {\it Suzaku} XIS instruments.
We used only XIS1 and XIS3 for image analysis,
since a part of XIS0 showed a known anomaly
(JX-ISAS-SUZAKU-MEMO-2010-01) and discriminated that region
in the data process before deriving spectra.
The vignetting has been corrected in each image
using {\tt xissim} \citep{ishisaki2007}.

One can see clear shell-like structures
in both images.
The shell has a rather clumpy and bulged structure.
The Western part of the X-ray shell has a smaller
curvature than the North-Western and Southern shell segments.
At the center of the remnant, emission from the point source
XMMU~J173203.3$-$344518 \citep{tian2008} is seen,
which from its soft spectrum and its position
has been classified as the central compact object (CCO)
\citep{acero2009,abramowski2011}.
\citet{tian2010} noted that the source spectrum is
also compatible with a magnetar hypothesis.
A claim of X-ray pulsation with a period of $\sim$1~s using
{\it XMM-Newton} data also led \citet{halpern2010}
to speculate that XMMU~J173203.3$-$344518 might be a magnetar candidate,
but a subsequent dedicated search for pulsations using  Chandra timing
observations did not confirm the signal \citep{halpern2010b}.

\begin{figure}
\epsscale{1.0}
\plotone{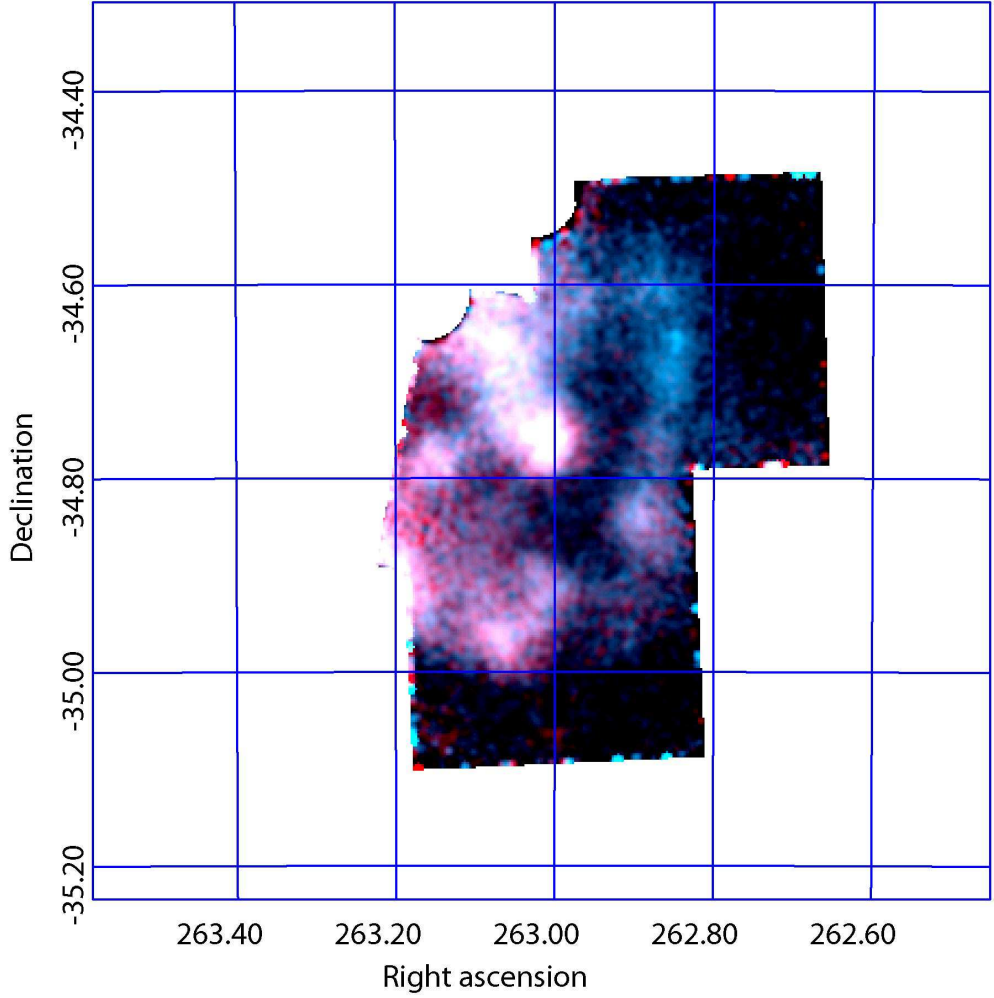}
\caption{
A false-color reprensentation of the HESS~J1731$-$347 
North-Western, Western, and Southern regions.
Red and blue correspond to 0.5--2.0~keV and 2.0--8.0~keV X-rays,
respectively.
The image is binned with 4~arcsec and smoothed with
a two-dimentional Gaussian of $\sigma=12$~arcsec.
}
\label{fig:image_color}
\end{figure}

In Fig.~\ref{fig:image_color}, a false-color image 
of the remnant is shown.
The image reveals that the North-Western part of this object
is characterized by
significantly harder
emission compared with 
the other regions.

\subsection{Multiwavelength comparison}

\begin{figure}
\epsscale{0.6}
\plotone{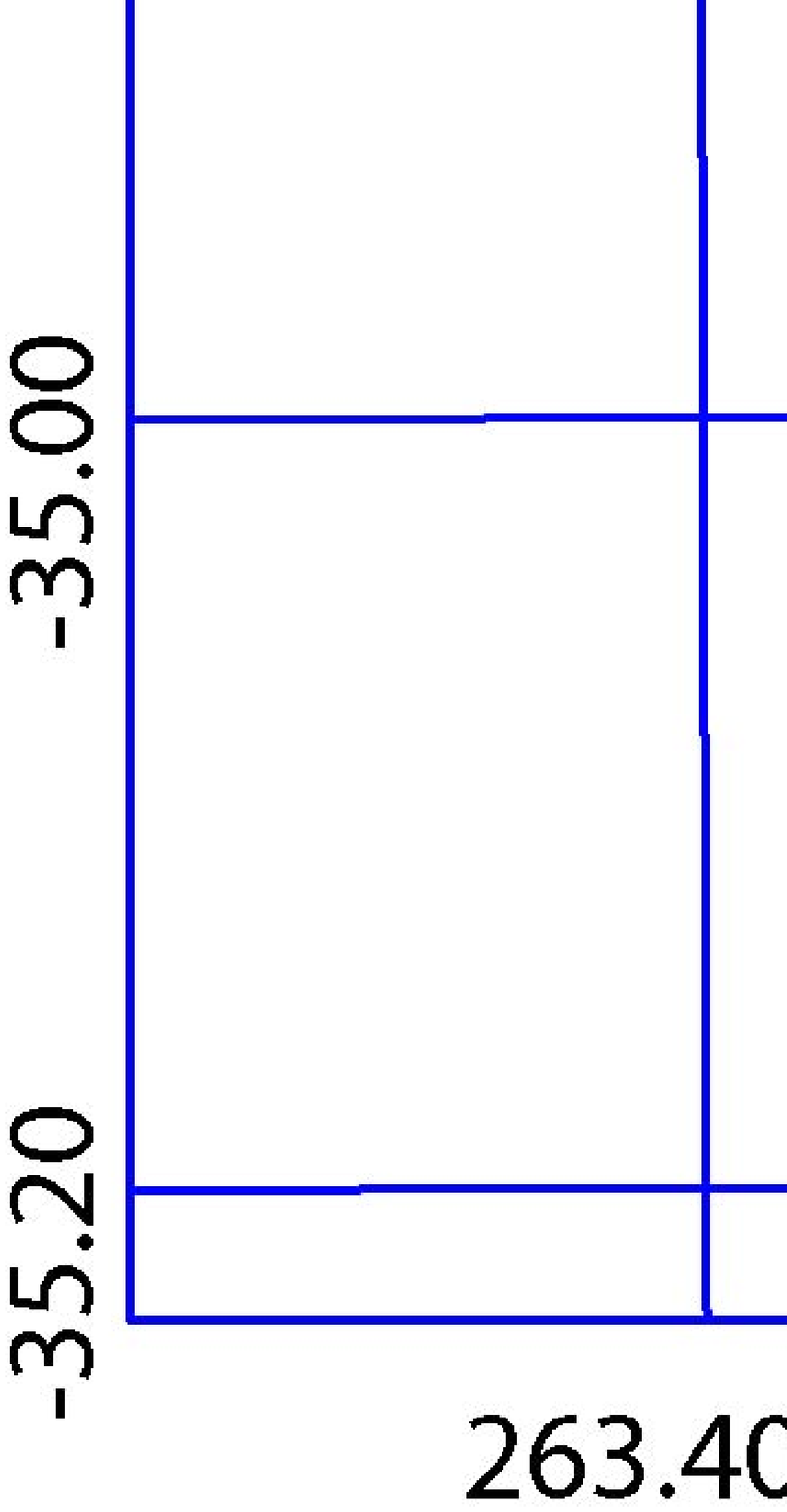}
\plotone{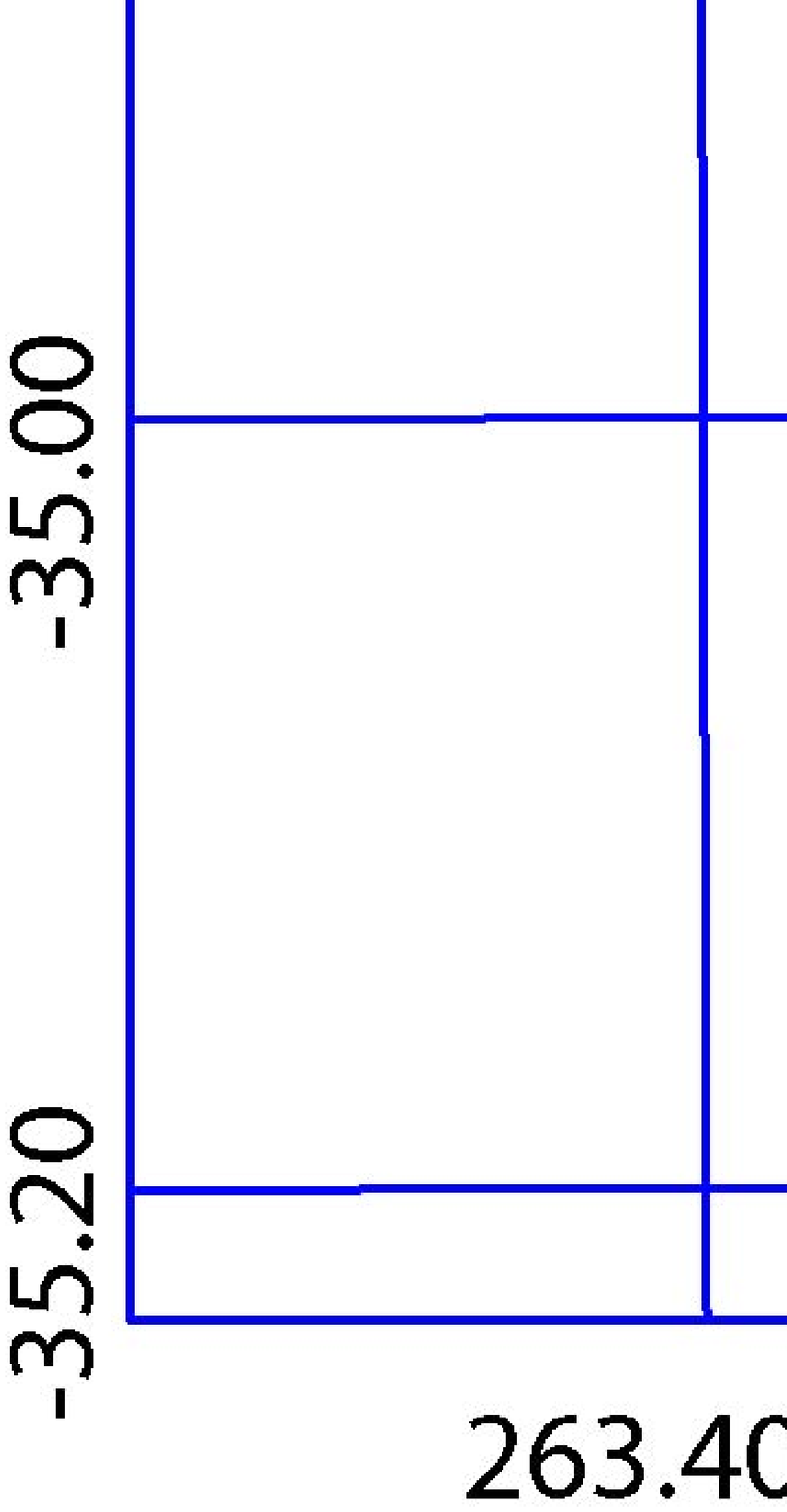}
\caption{
Multiwavelength view of the HESS~J1731$-$347 region.
The color images show the {\it Suzaku} data in a false-color
representation with the same color scaling as in Fig.~\ref{fig:image_color},
whereas the contours are derived from the ATCA radio map at 1.4~GHz
(top)
and from the H.E.S.S. VHE gamma-ray excess count map
\citep{abramowski2011} (bottom).
The green circle denotes the best-fit model of the VHE shell,
which has a radius of 0.27~deg. and an unresolved shell width.
}
\label{fig:multiwavelength}
\end{figure}

This SNR has a shell-like structure
not only in X-rays but also in radio and VHE gamma-rays
\citep{tian2008,abramowski2011}.
The top panel of Figure~\ref{fig:multiwavelength} shows
a comparison of the {\it Suzaku} data with a radio map at 1.4~GHz
from the ATCA south Galactic Plane survey \citep{haverkorn2006}.
Both images show clumpy shell structures,
and an overall good agreement with each other.
The bottom panel of Fig.~\ref{fig:multiwavelength}
shows a comparison of the {\it Suzaku} data with the VHE gamma-ray map
derived from H.E.S.S. observations \citep{abramowski2011}.
At first glance, it looks as if there are differences in the extent of
the VHE and X-ray emissions towards the North-West and West of the source,
respectively.
Those differences might however be attributed to
the coarser angular resolution
of the VHE data (0.06~deg., 68\% containment radius) compared with the 
{\it Suzaku} X-ray data
\citep[half power diameter of $\sim$2~arcmin,][]{mitsuda2007}.
We refer to the result of \citet{abramowski2011} that
the radio and VHE radio profiles of the entire remnant match well
when the radio data are blurred to VHE resolution.

\begin{figure}
\epsscale{0.5}
\plotone{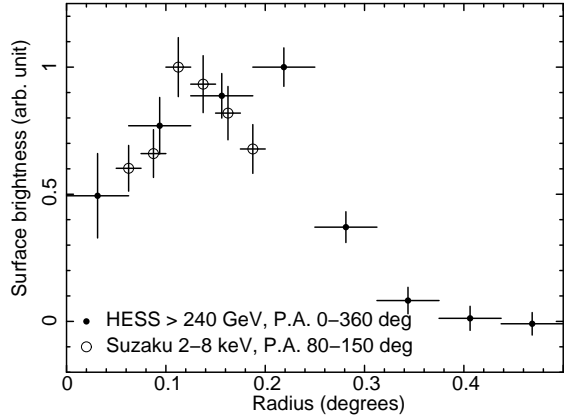}
\caption{
VHE $\gamma$-ray (filled circle) and 2--8~keV (open circle)
radial profiles.
Both profiles are centered on the central compact object,
(RA, Dec.) = (263.0125, -34.755).
The central part was cut in the X-ray data
due to the contamination from the central source.
Note that the VHE $\gamma$-ray profile is taken from 
the entire remnant, whereas the X-ray one is from
the Western part of the shell,
using position angle (P.A.) 80$^\circ$--150$^\circ$
(counted clockwize starting from North).}
\label{fig:profile}
\end{figure}

To illustrate the potential difference
between the X-ray and the VHE radial profiles,
in Fig.~\ref{fig:profile} the 2--8~keV profile for 
a wedge towards the west is compared to 
the VHE profile derived from the entire remnant.
Besides the difference in angular resolution,
it is also not yet feasible with the current VHE statistics
to restrict the VHE data to a similarly small region.
Further VHE observations
using e.g. the planned Cherenkov Telescope Array CTA \citep{actis2011}
are needed to provide sufficient VHE photon statistics
to permit morphological comparisons on such small angular scales.

\subsection{Spectra below 10~keV}

\subsubsection{Diffuse emission}

\begin{figure}
\epsscale{0.8}
\plotone{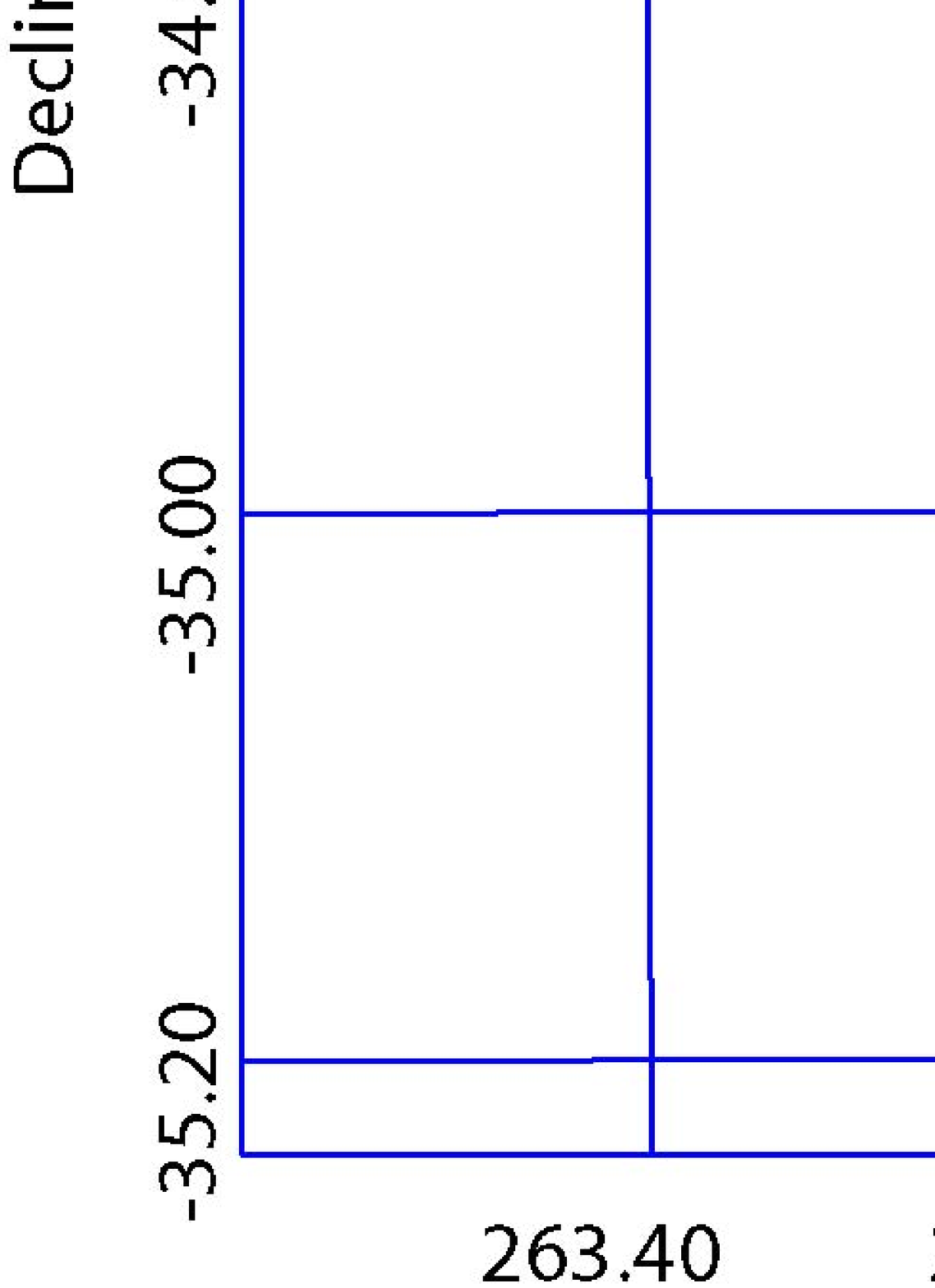}
\plotone{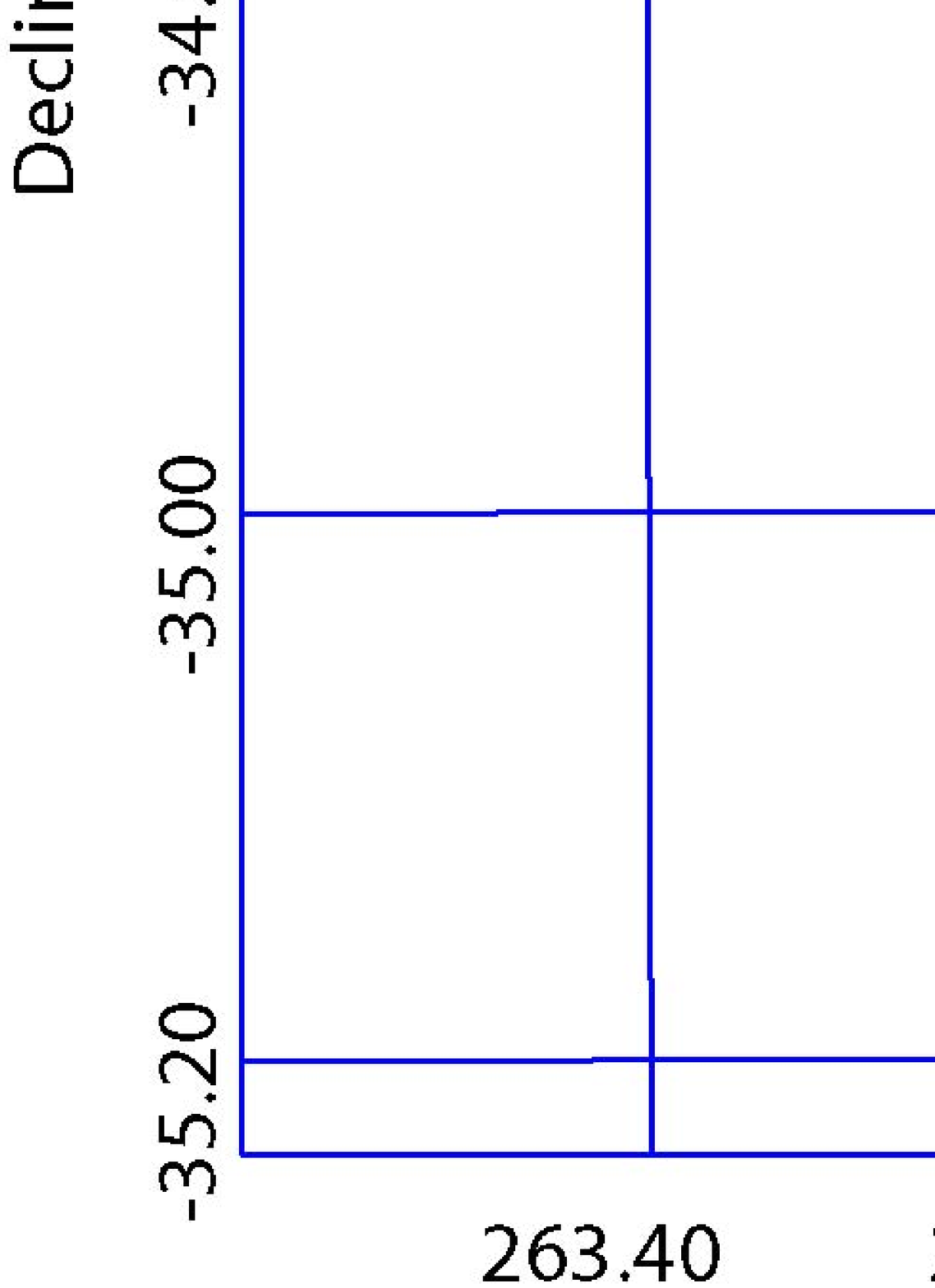}
\caption{
{\it Suzaku} XIS intensity maps of the North-Western, Western,
and Southern regions of HESS~J1731$-$347
in the 0.5--2.0~keV (top) and 2.0--8.0~keV (bottom) bands.
The image is binned with 4~arcsec and smoothed with
a two dimentional Gaussian of $\sigma=12$~arcsec.
Solid boxes in the top panel
represent the source (thick) and background (thin) regions
used for the spectral study.
The unit of the color bars in the panels is 
photons~s$^{-1}$pixel$^{-1}$.
}
\label{fig:images}
\end{figure}

In order to perform a spatially-resolved spectral study below 10~keV
with XIS,
we divided the X-ray map into several regions,
as shown in Figure~\ref{fig:images}.
The background regions were taken from source-free regions
of the same observations.

\begin{figure}
\epsscale{0.25}
\plotone{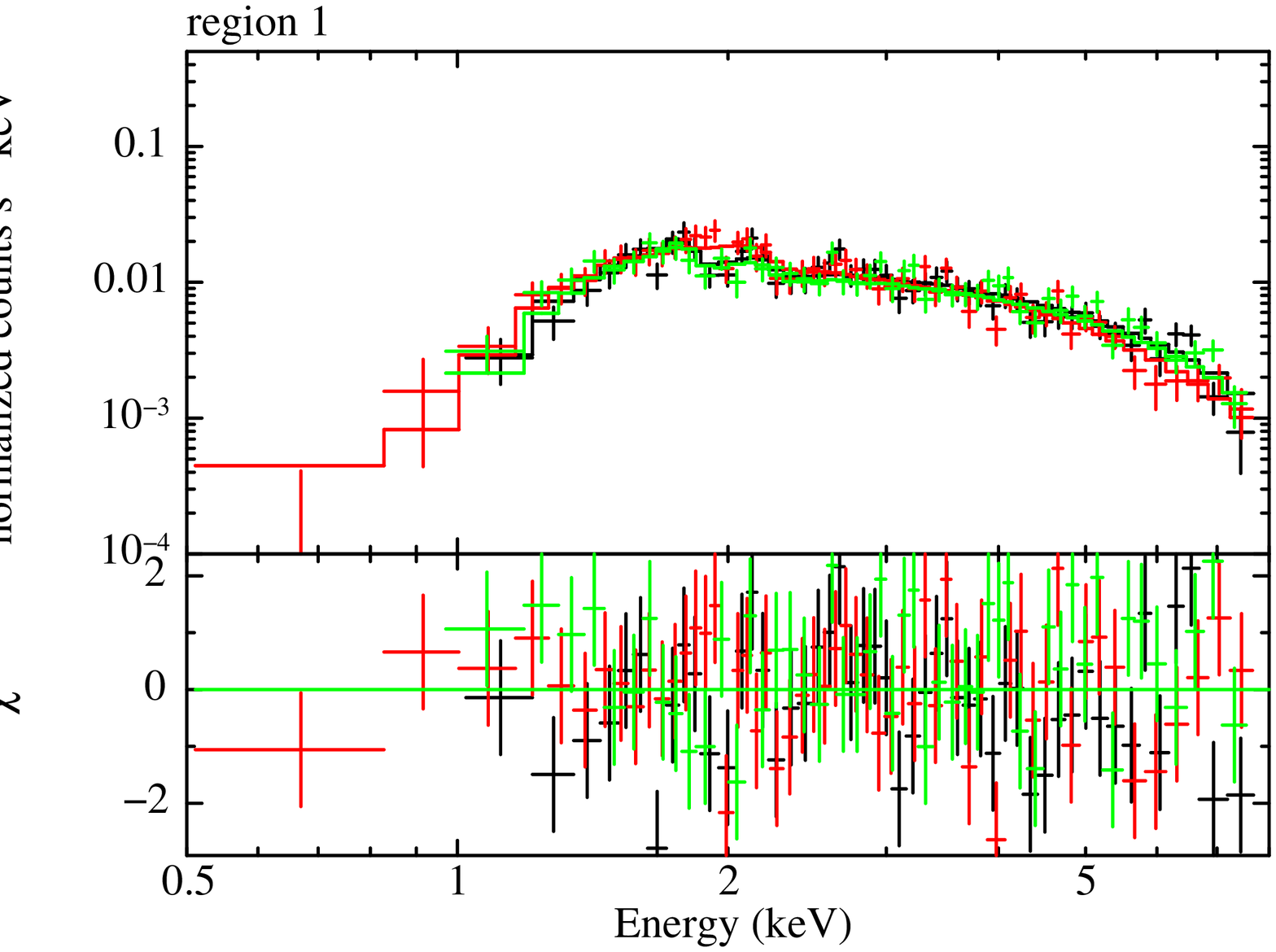}
\plotone{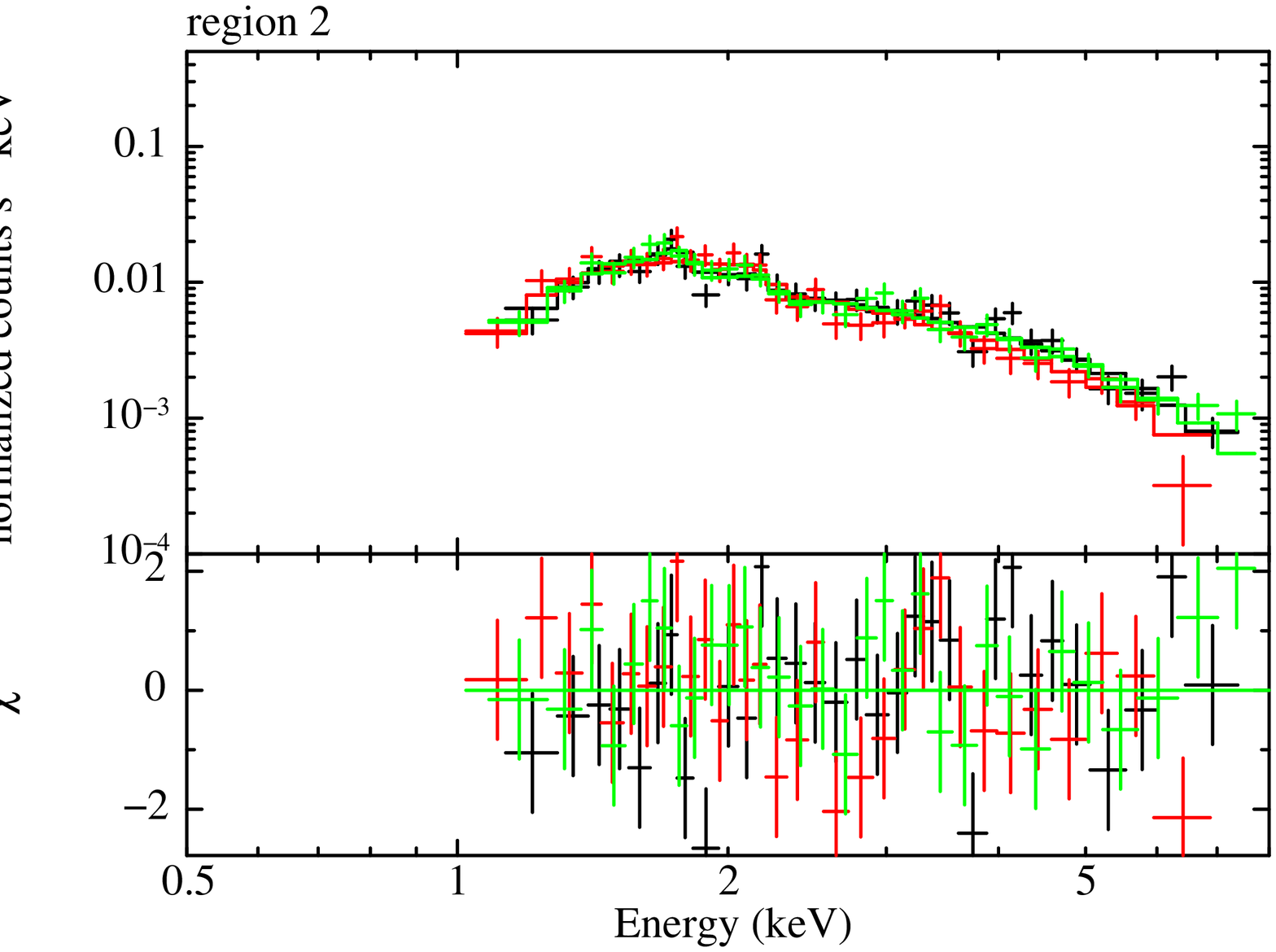}
\plotone{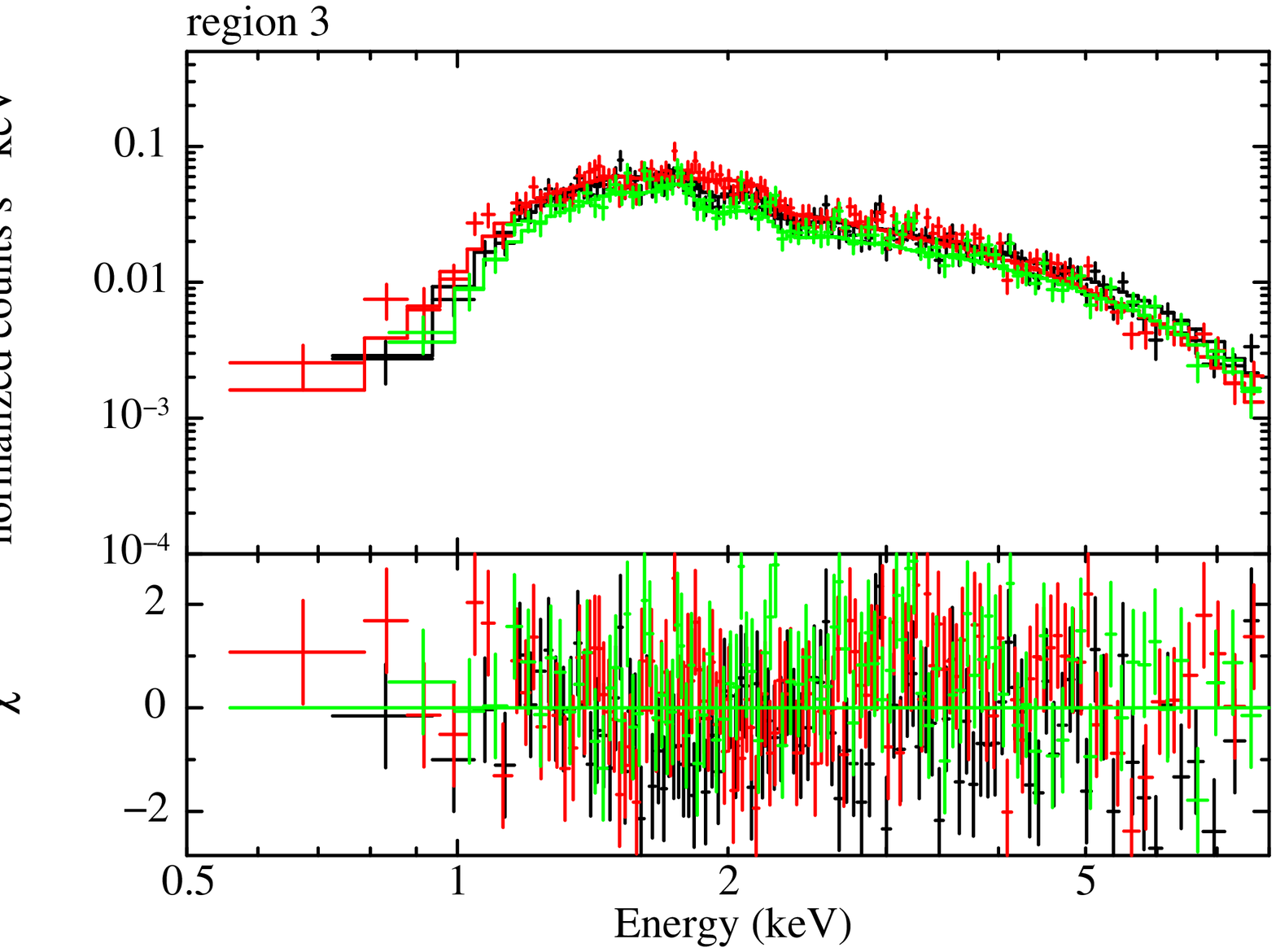}
\plotone{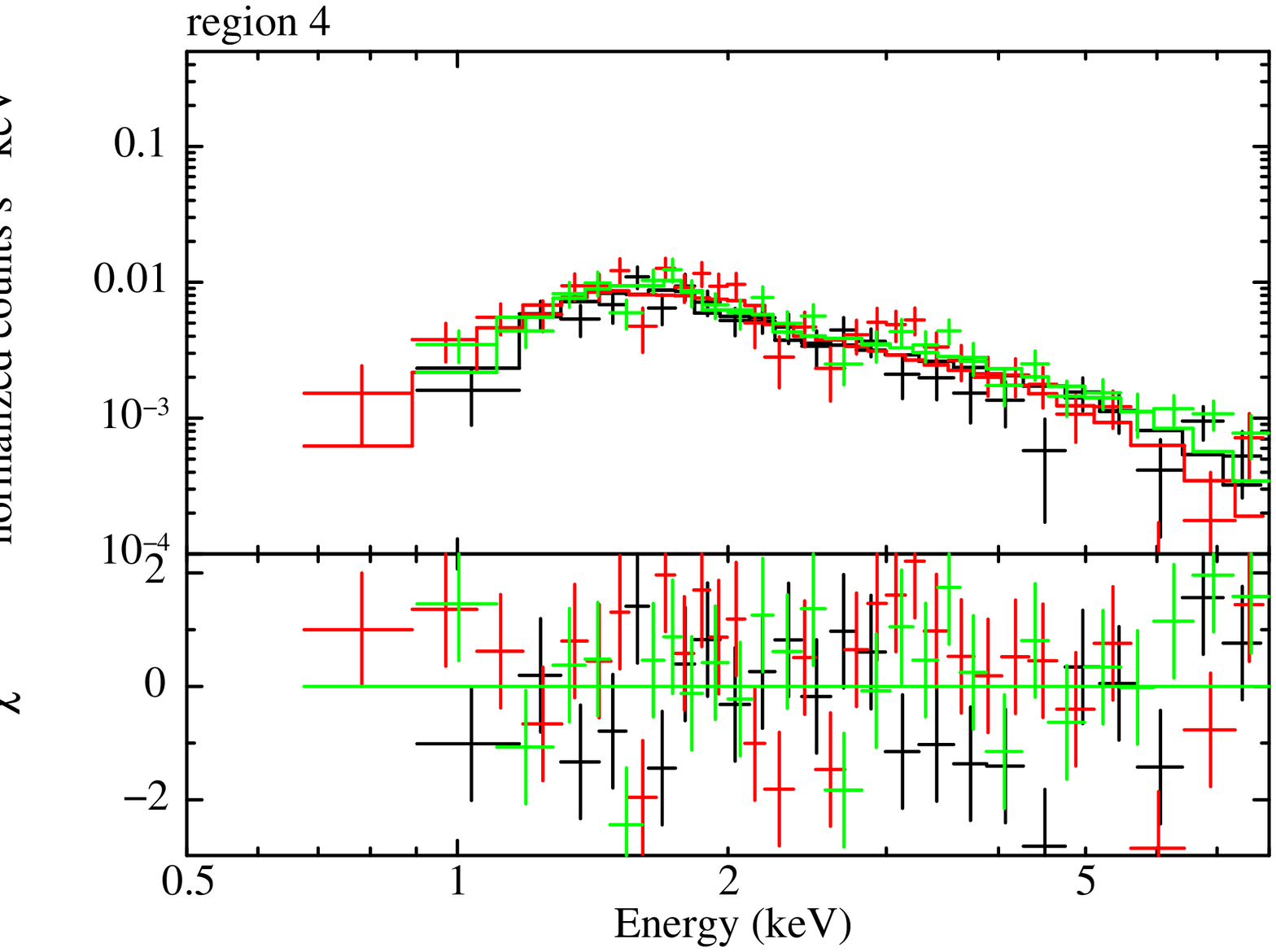}
\plotone{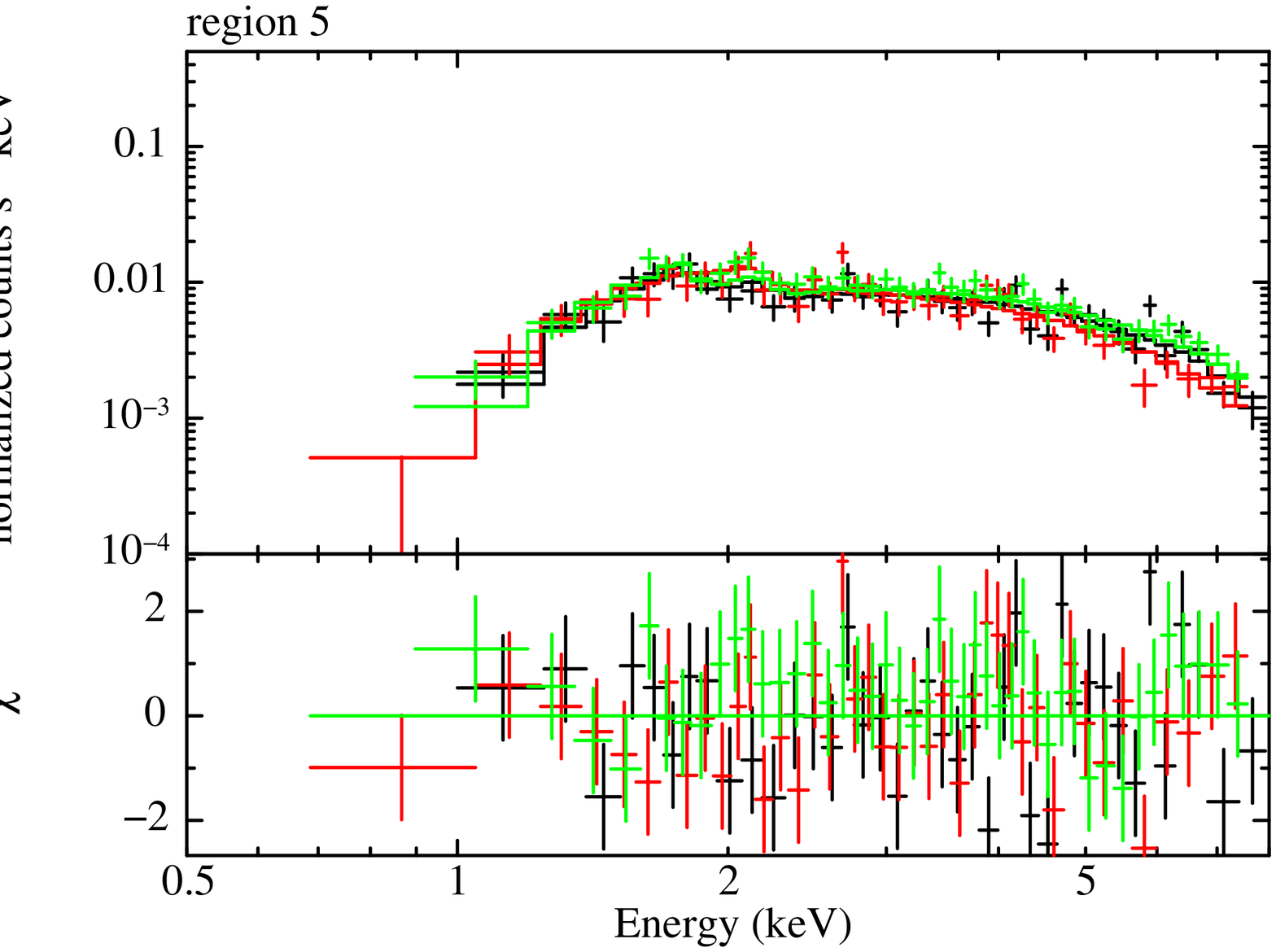}
\plotone{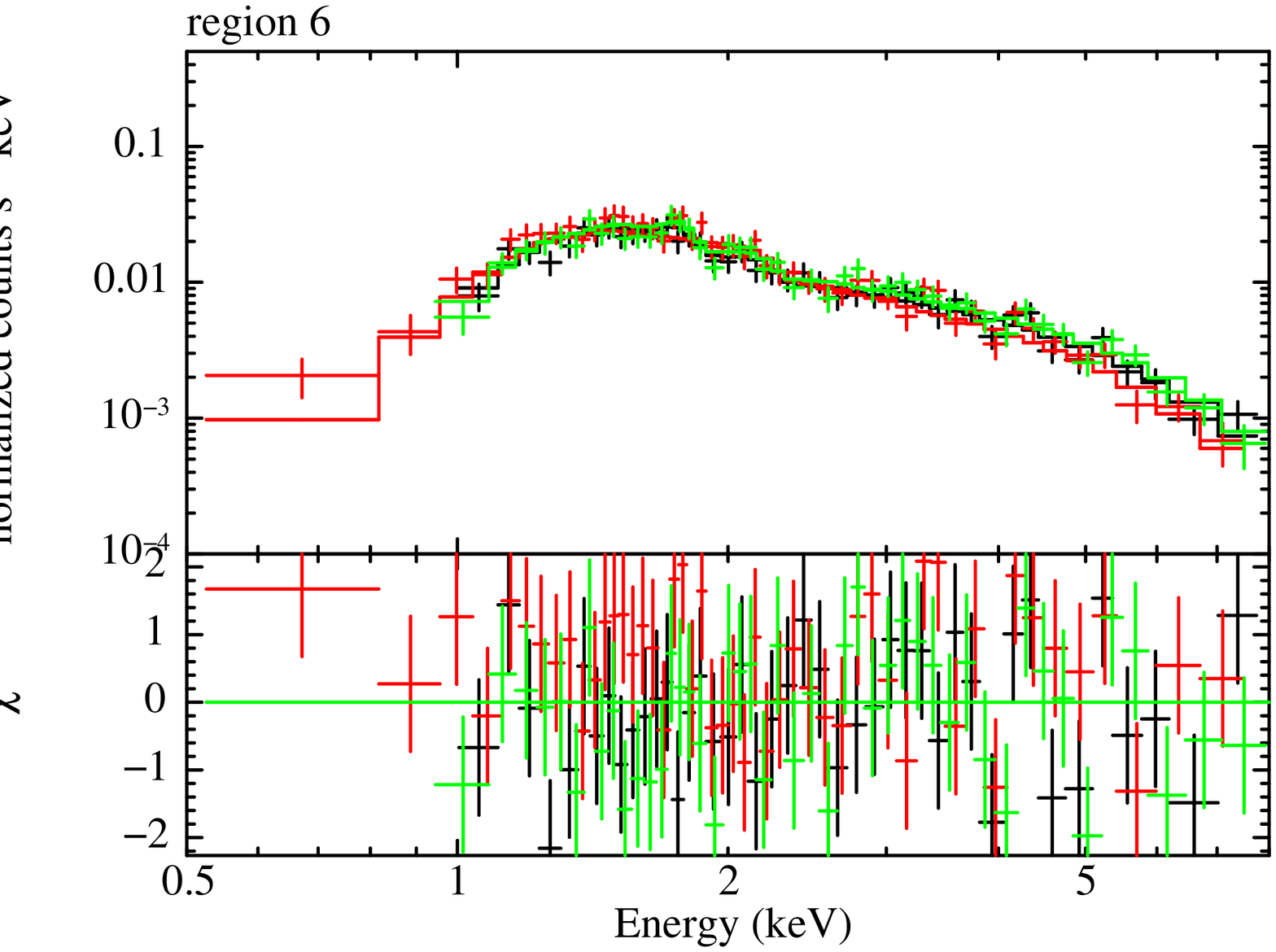}
\plotone{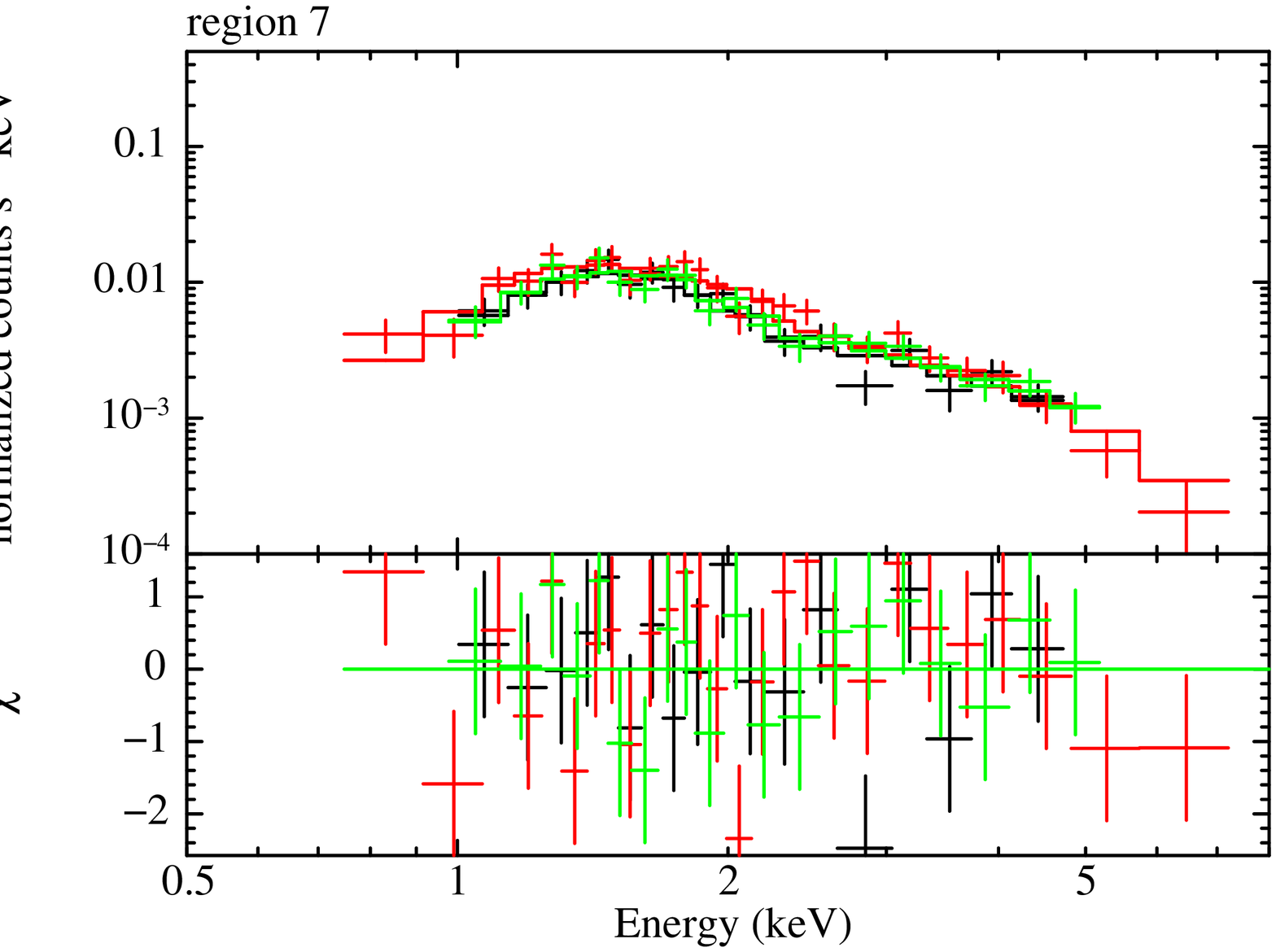}
\plotone{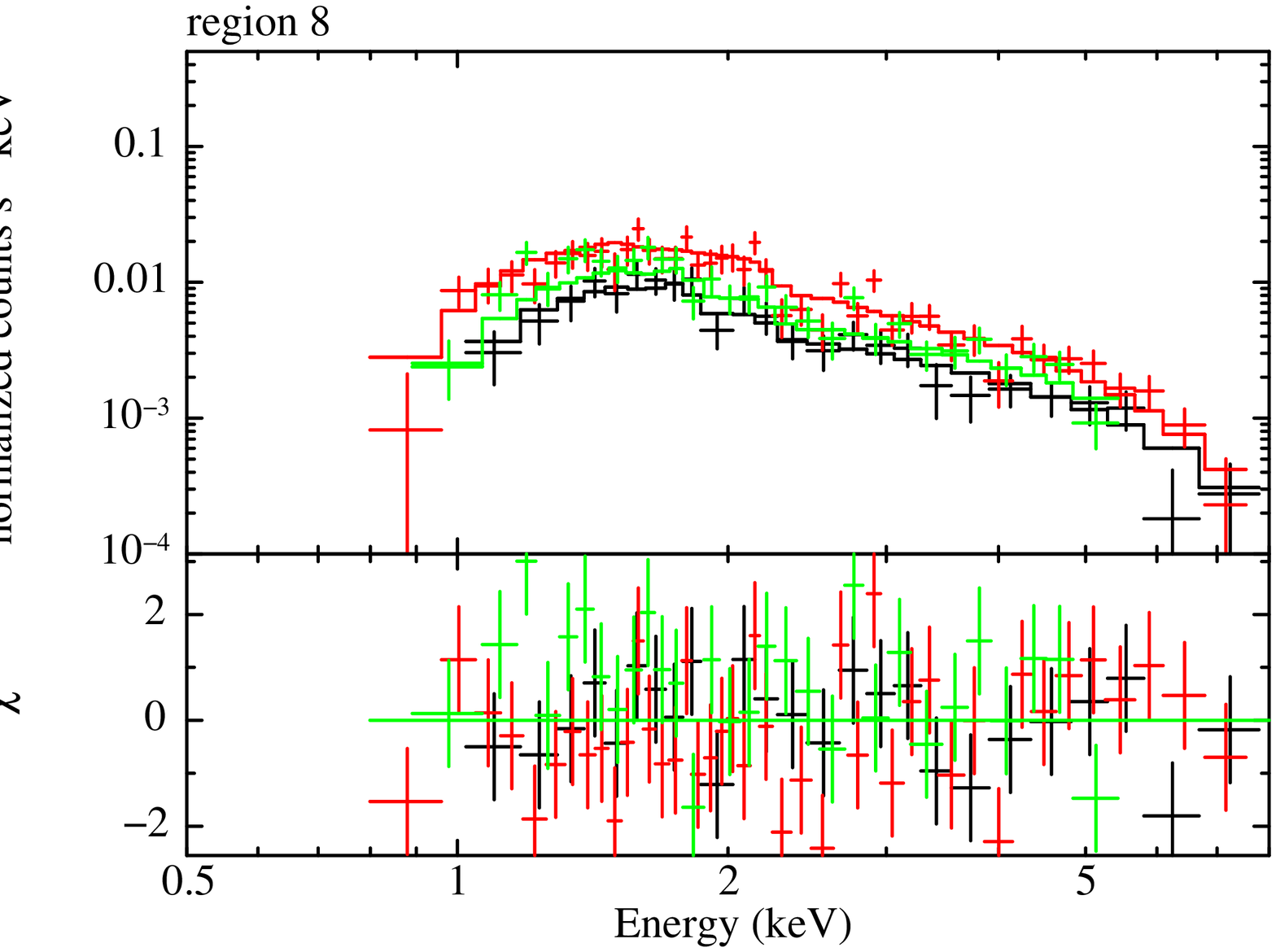}
\plotone{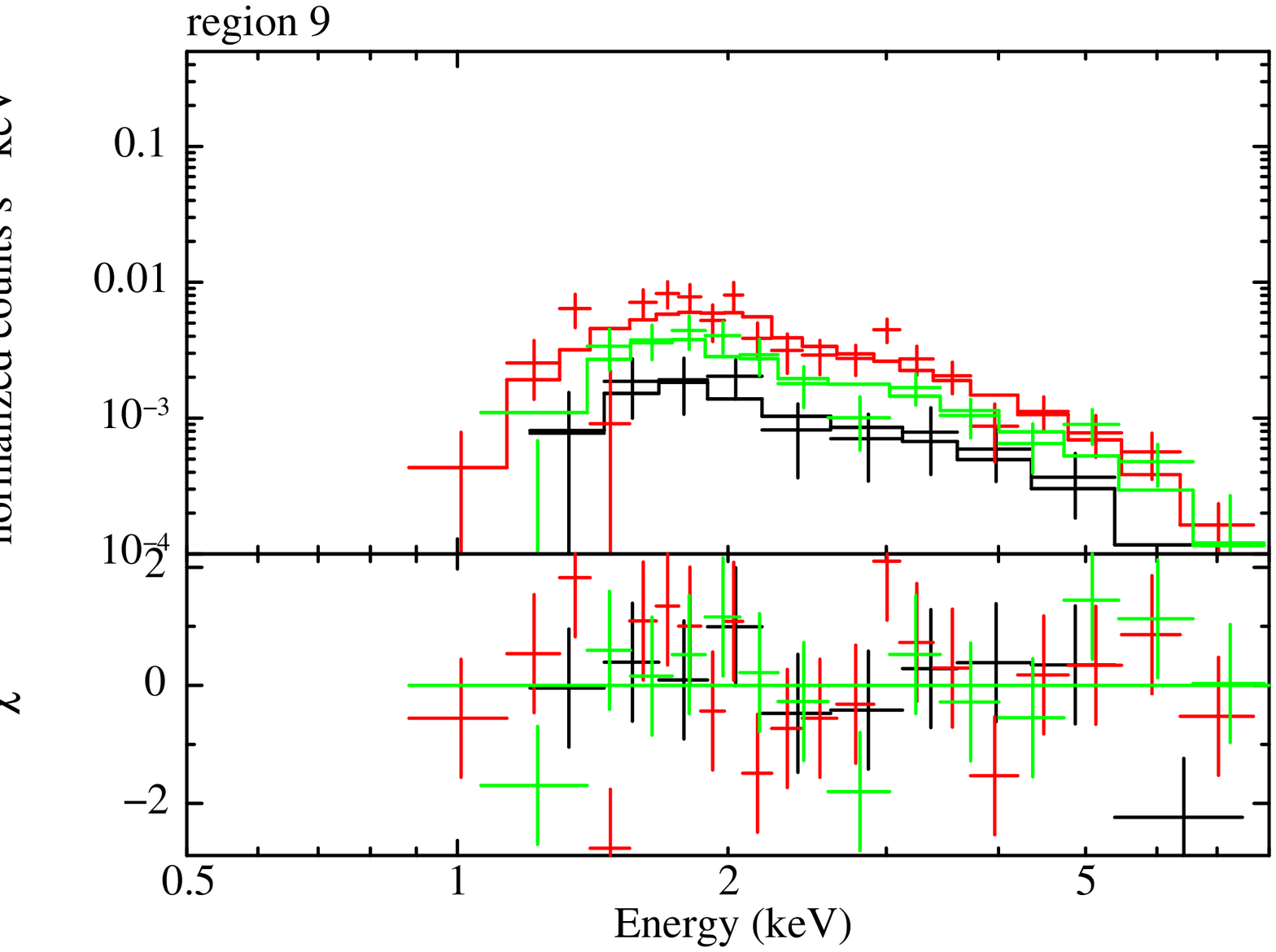}
\plotone{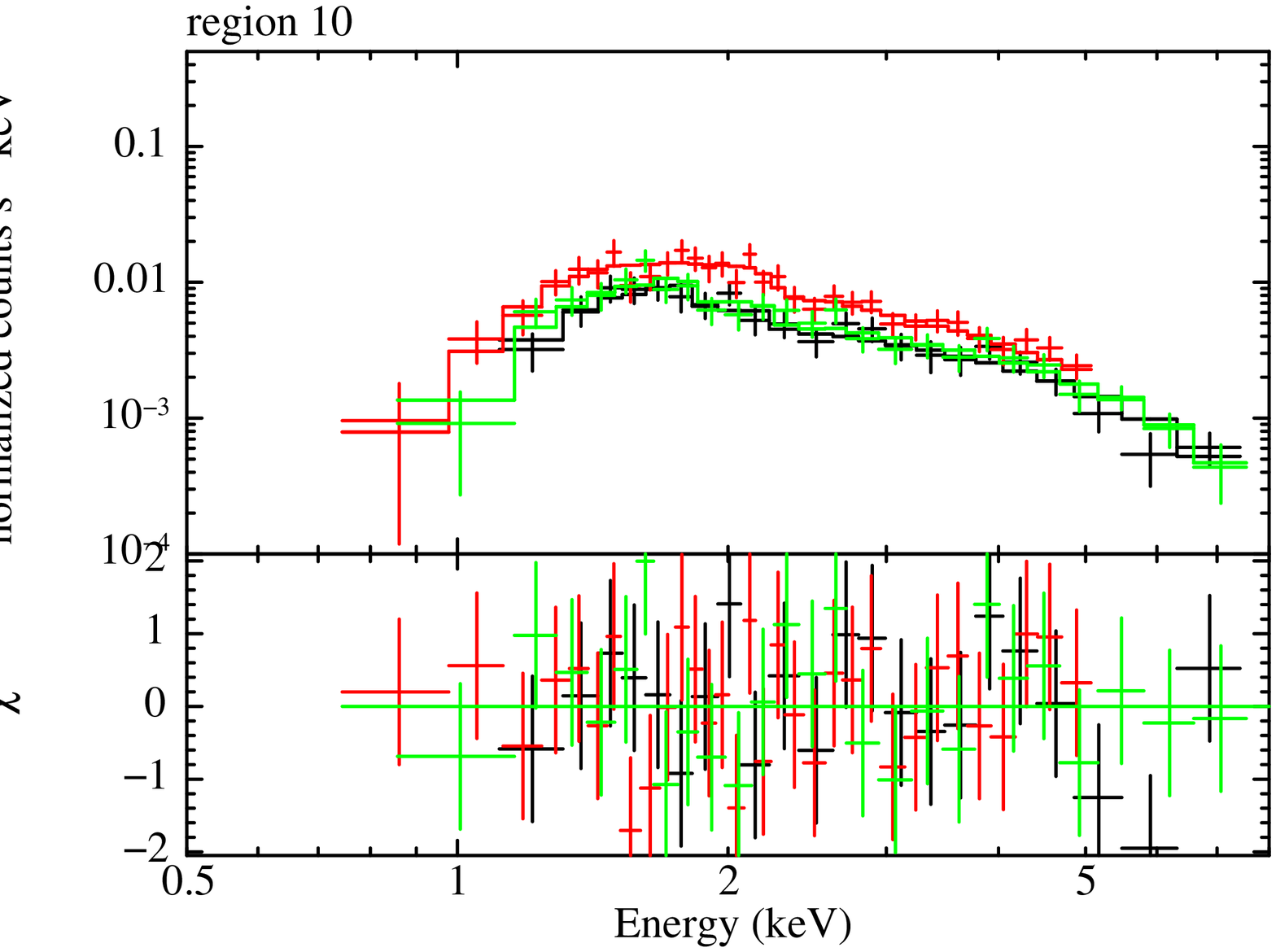}
\plotone{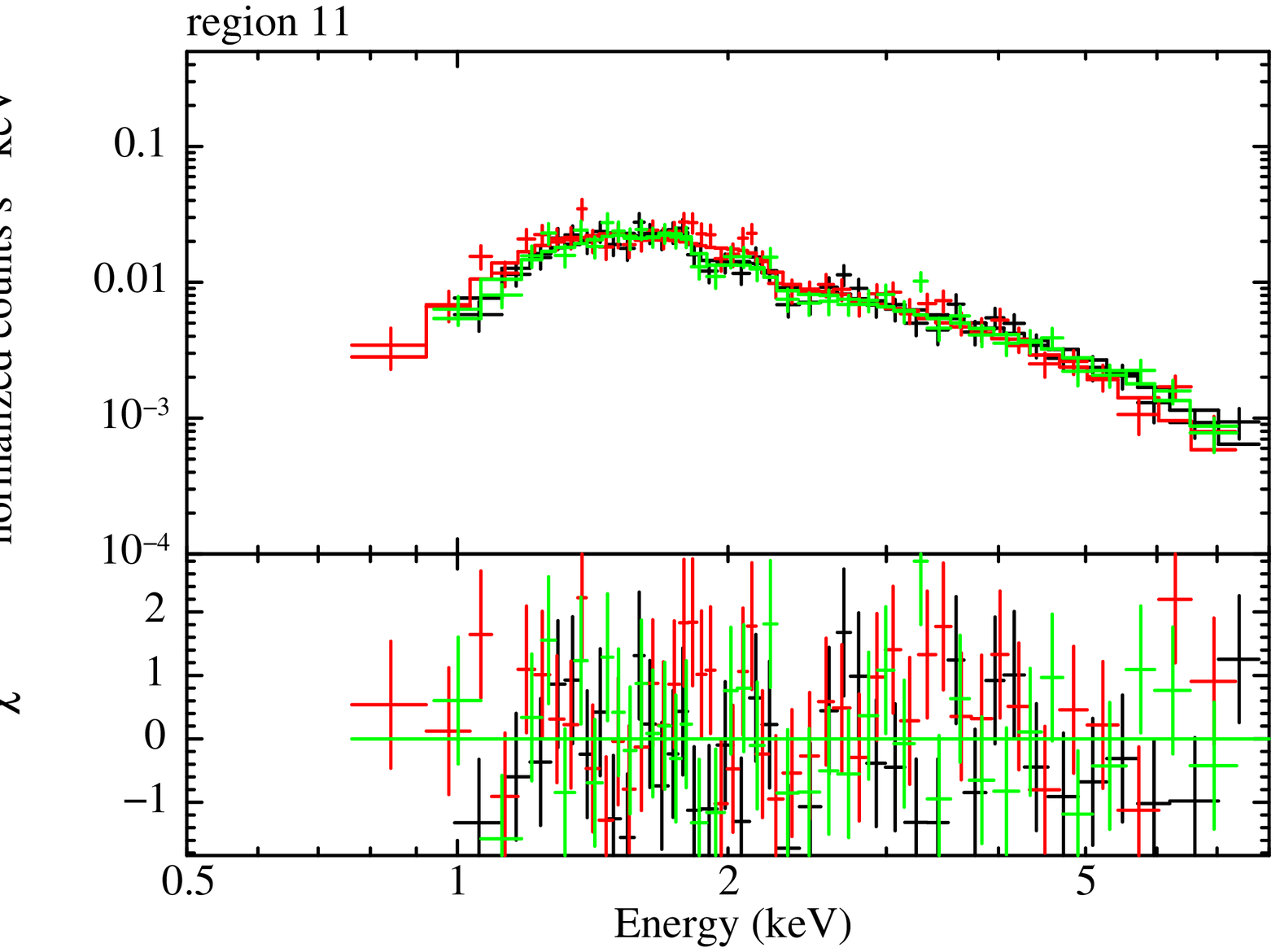}
\plotone{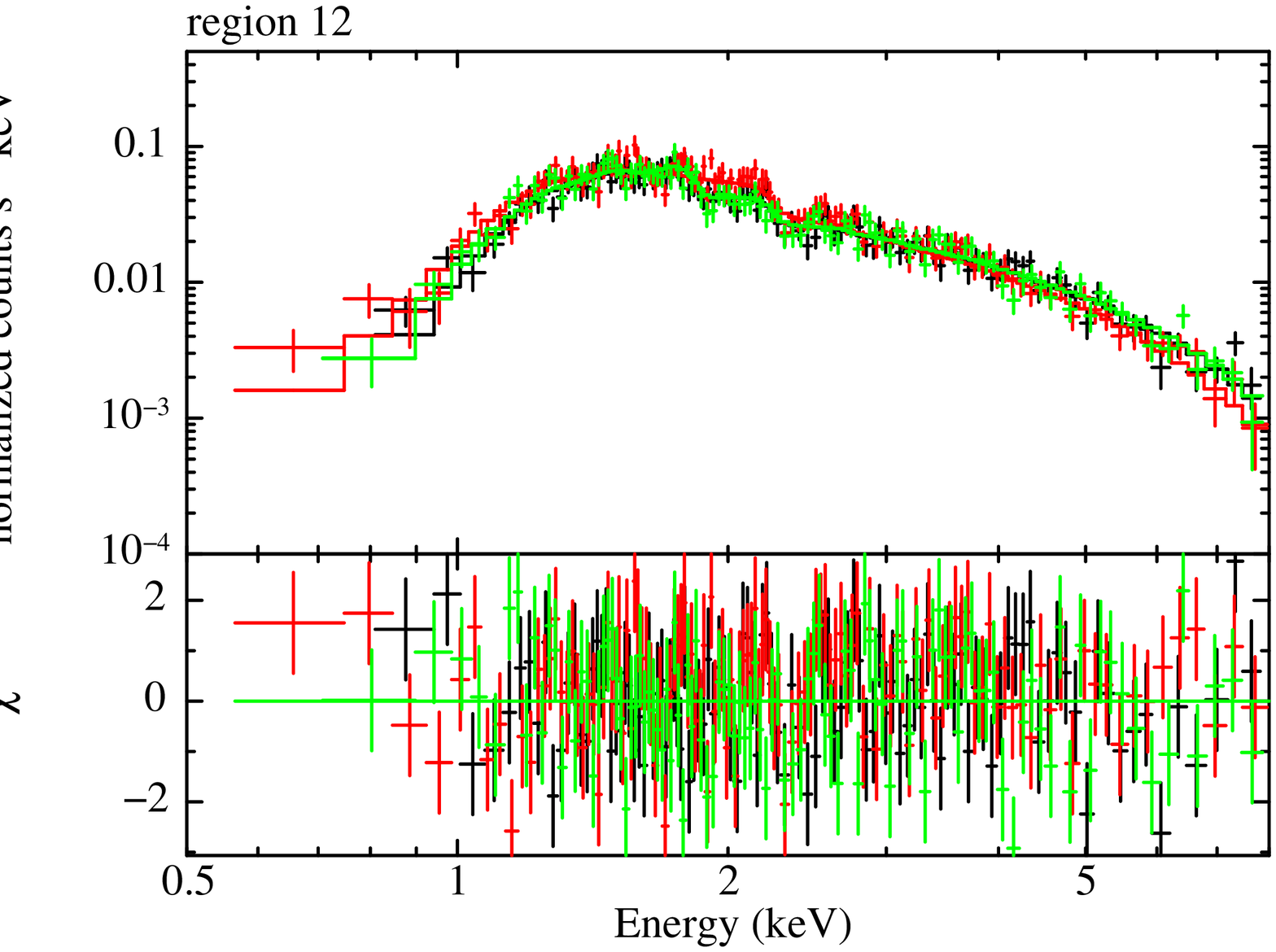}
\plotone{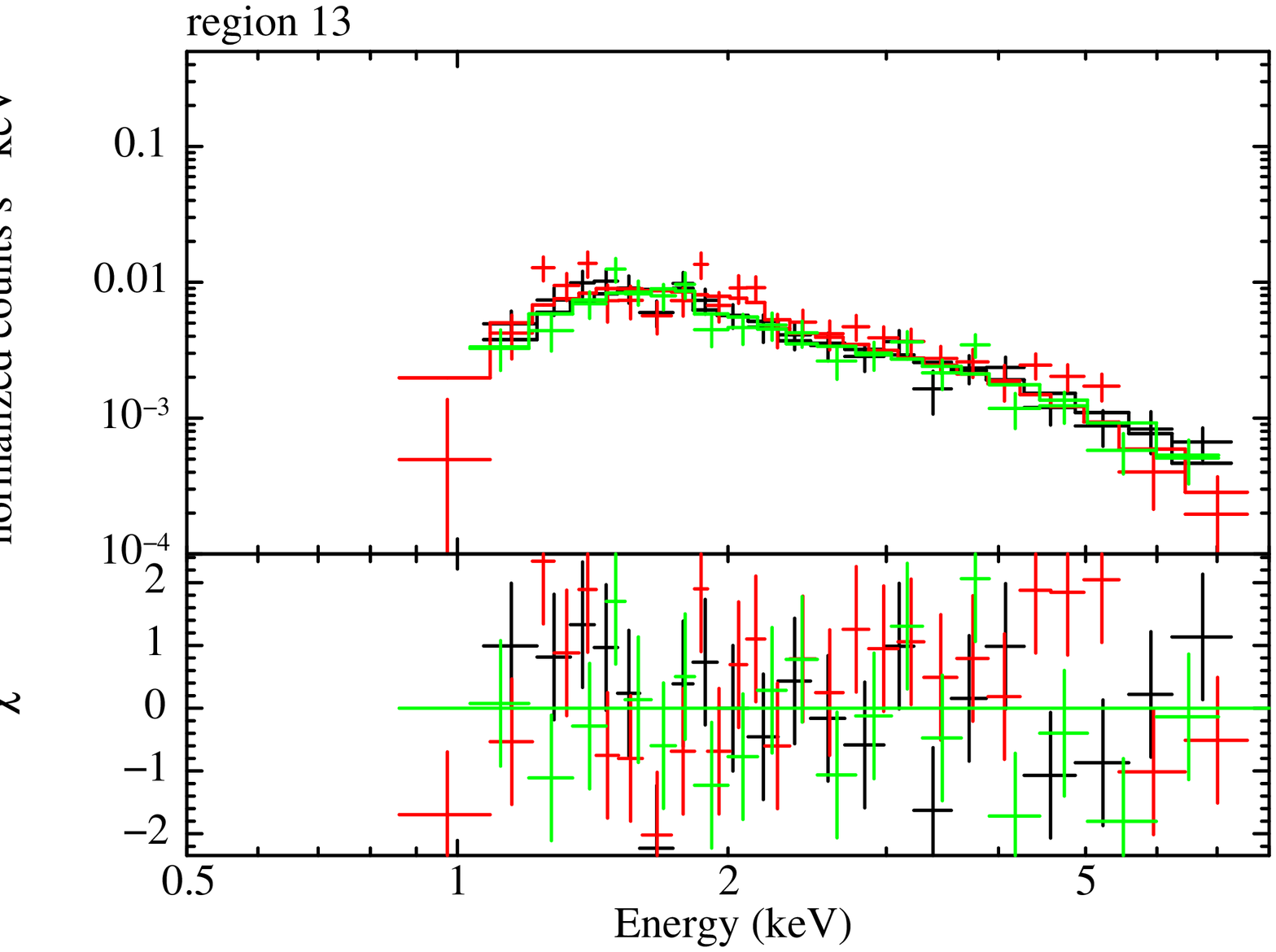}
\plotone{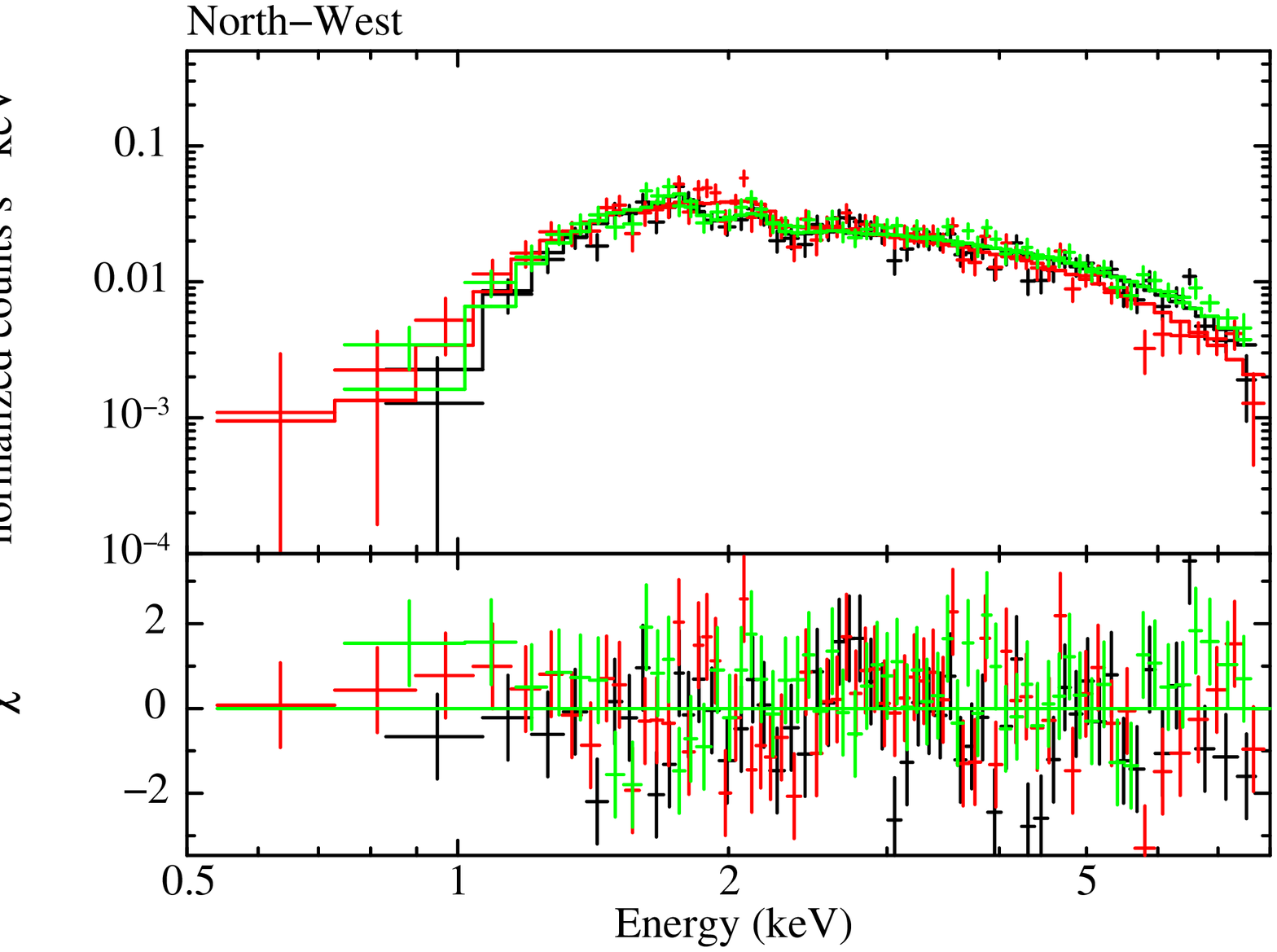}
\plotone{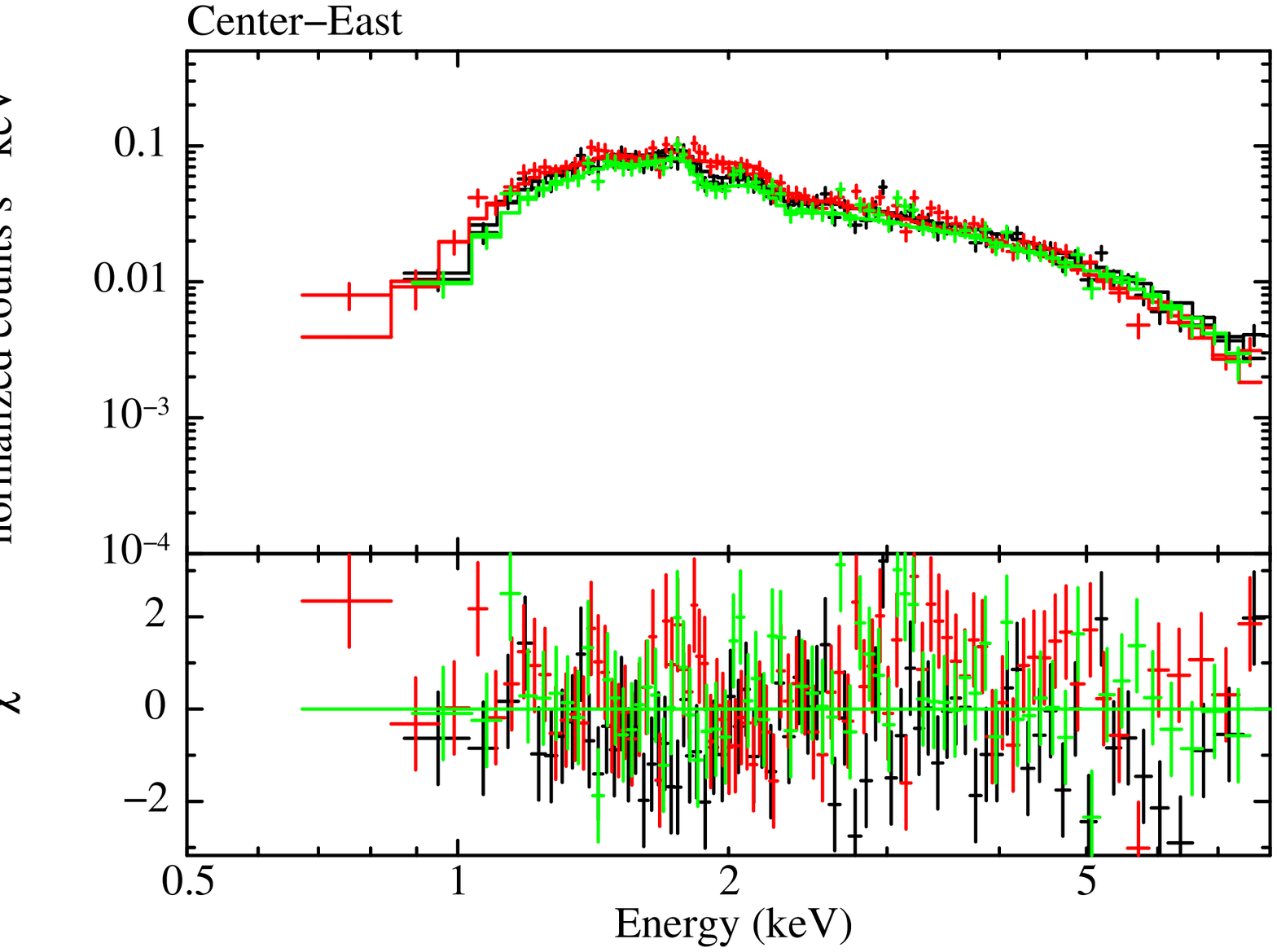}
\plotone{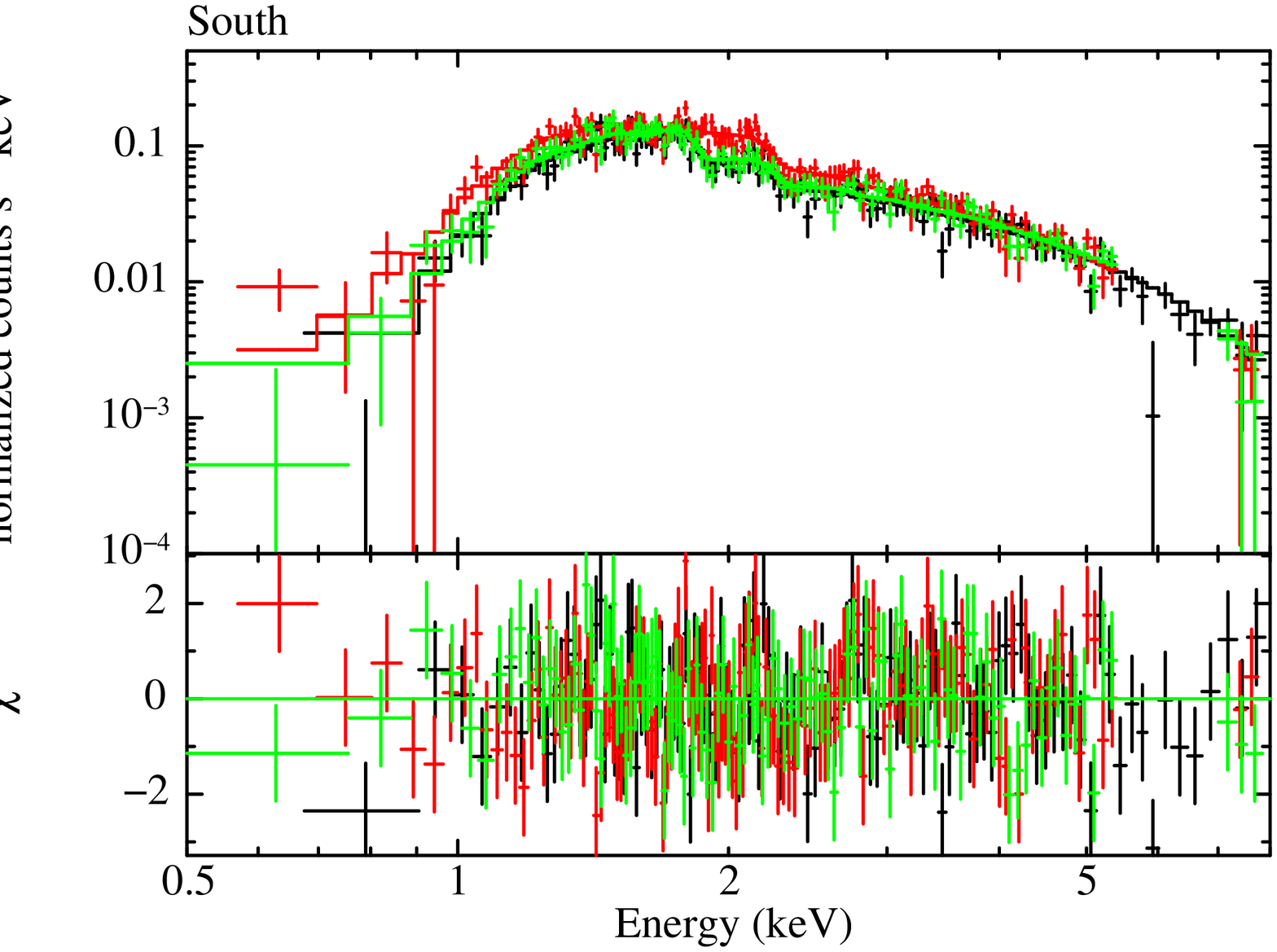}
\caption{XIS spectra of each region
(for the region definition see Fig.~\ref{fig:images})
and combined one (see text).
The color represents XIS0 (black), XIS1 (red), and XIS3 (green).
The solid lines represent the best-fit absorbed power-law model.
The lower panels show the residuals from the best-fit models,
respectively.
}
\label{fig:spectra}
\end{figure}

As can be seen in Figure~\ref{fig:spectra},
all background-subtracted spectra are hard and
show no evidence of line emission.
We thus fitted the spectra with an absorbed power-law model.
The absorption model includes the cross sections of
\citet{balcinska-church1992} with solar abundance \citep{anders1989}.
Auxiliary files for the effective area
were produced with {\tt xissimarfgen} \citep{ishisaki2007},
using the observed images of the source regions
as photon distribution data.
For each region, the spectra of all available XIS detectors
were fitted simultaneously.
For some regions, less than three XIS spectra are available
because the regions overlap with the damaged part of XIS0 or
with calibration sources.
Because of this,
we allowed for a difference of the normalization between different XIS spectra,
and adopted the best-fit normalization derived from XIS3 as final result,
since any remaining contamination from calibration sources is smallest
in XIS3 for the regions analysed.
Figure~\ref{fig:spectra} and Table~\ref{tab:spectra} show
our best-fit models and parameters with 90\% errors.
In each case the power-law fit was statistically acceptable.

\begin{figure}
\epsscale{0.45}
\plotone{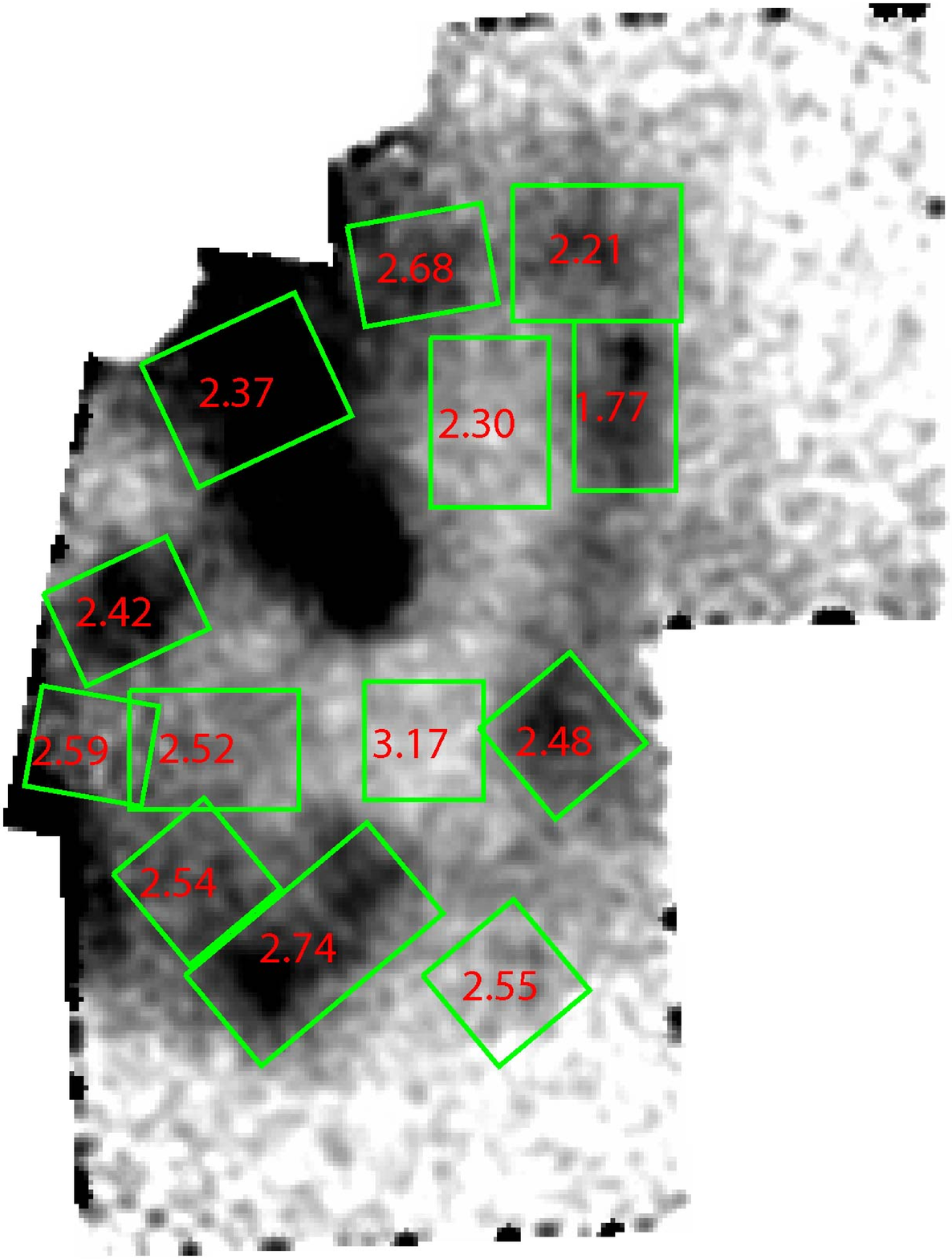}
\plotone{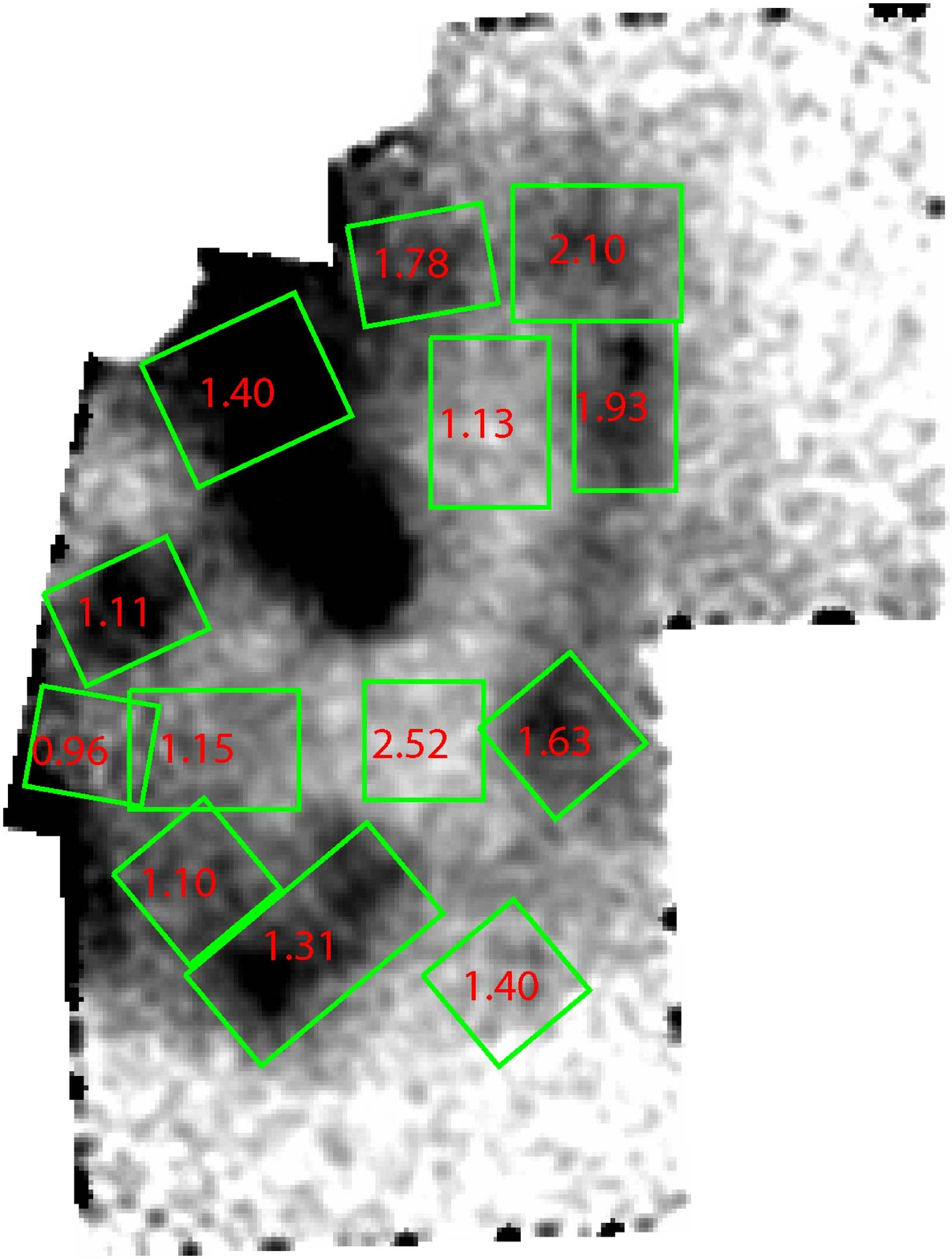}
\caption{
Best-fit photon index (left) and absorption column density (right)
in each region.
}
\label{fig:spectral_change}
\end{figure}

Figure~\ref{fig:spectral_change} shows the best-fit photon index (left)
and column density (right)
for all analysed regions of the SNR.
The North-Western part of the shell has
a significantly harder and more absorbed spectrum,
compared with the rest of the covered remnant.
On the other hand, differences between other individual regions
are not significant when statistical and systematic errors
(at low surface-brightness regions, see below) are taken into account.

In order to better constrain the spectral shape of the non-thermal emission,
we combined several adjacent regions to three cardinal areas:
regions 1, 4, and 5 to the North-West,
regions 3 and 6 to the Center-East,
and regions 8--13 to the South,
respectively.
All the three spectra were again fitted well
with an absorbed power-law model
without any spectral break.
The spectra with the best-fit models and their best-fit parameters are shown in
Fig.~\ref{fig:spectra} and Tab.~\ref{tab:spectra}, respectively.
As already seen from the individual spectra,
also in the combined spectra
there is a significant change in photon index across the remnant,
with the hardest spectrum in the North-West,
where also the absorption column density is significantly larger 
than in the other two areas.

As illustrated in Fig.~\ref{fig:images},
we used two apparently source-free regions
to estimate the average background of the entire source region.
Due to the position dependence of
the Galactic Ridge X-Ray Emission (GRXE),
that can be $\sim$20\% contribution in this region \citep{uchiyama2010},
the spectral parameters of low surface brightness regions might be
misestimated.
In order to quantify this systematic error,
we artificially varied the background by $\pm$20\%
in the fit of region 9,
which has the lowest surface brightness of all regions analyzed.
Indeed,
the systematic error in this region is
comparable with the statistical error:
$\Gamma$ = 3.2 (+0.5$-$0.5)$_{\rm stat}$(+0.7$-$0.4)$_{\rm sys}$
and
$N_{\rm H}$ = (2.5 (+0.7$-$0.6)$_{\rm stat}$(+0.9$-$0.5)$_{\rm sys}$)
$\times 10^{22}~{\rm cm}^2$.
For the high surface brightness regions
($> 0.8\times 10^{-13}$~erg~cm$^{-2}$s$^{-1}$arcmin$^{-2}$,
see Table~\ref{tab:spectra})
which are relevant for the evaluation of the spectral and 
absorption trends,
the systematic errors are below the 90\% statistical errors.
Therefore, while the significance of the observed absorption
and photon index trends indeed weakens,
they are nevertheless significantly observed.

\begin{figure}
\epsscale{0.45}
\plotone{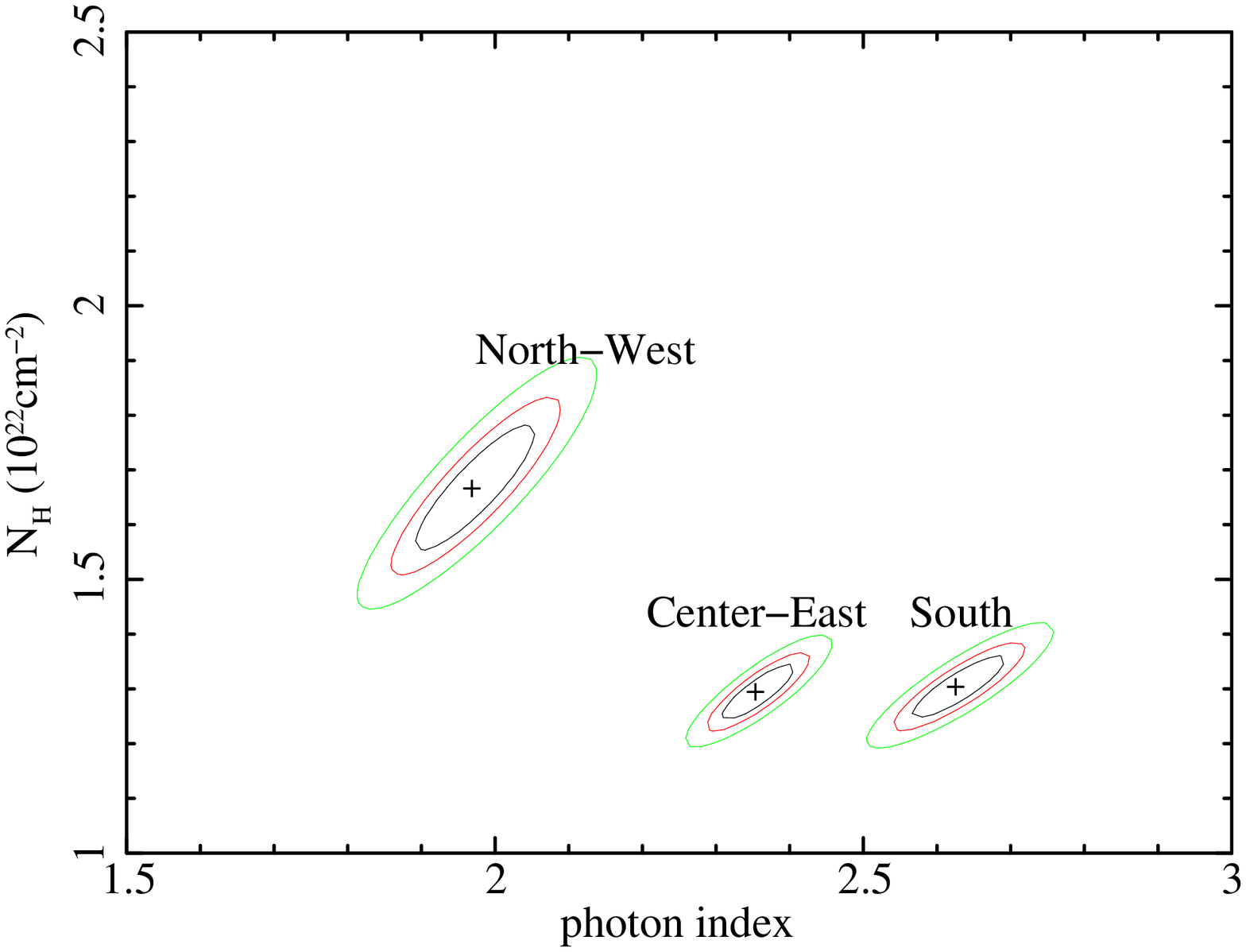}
\plotone{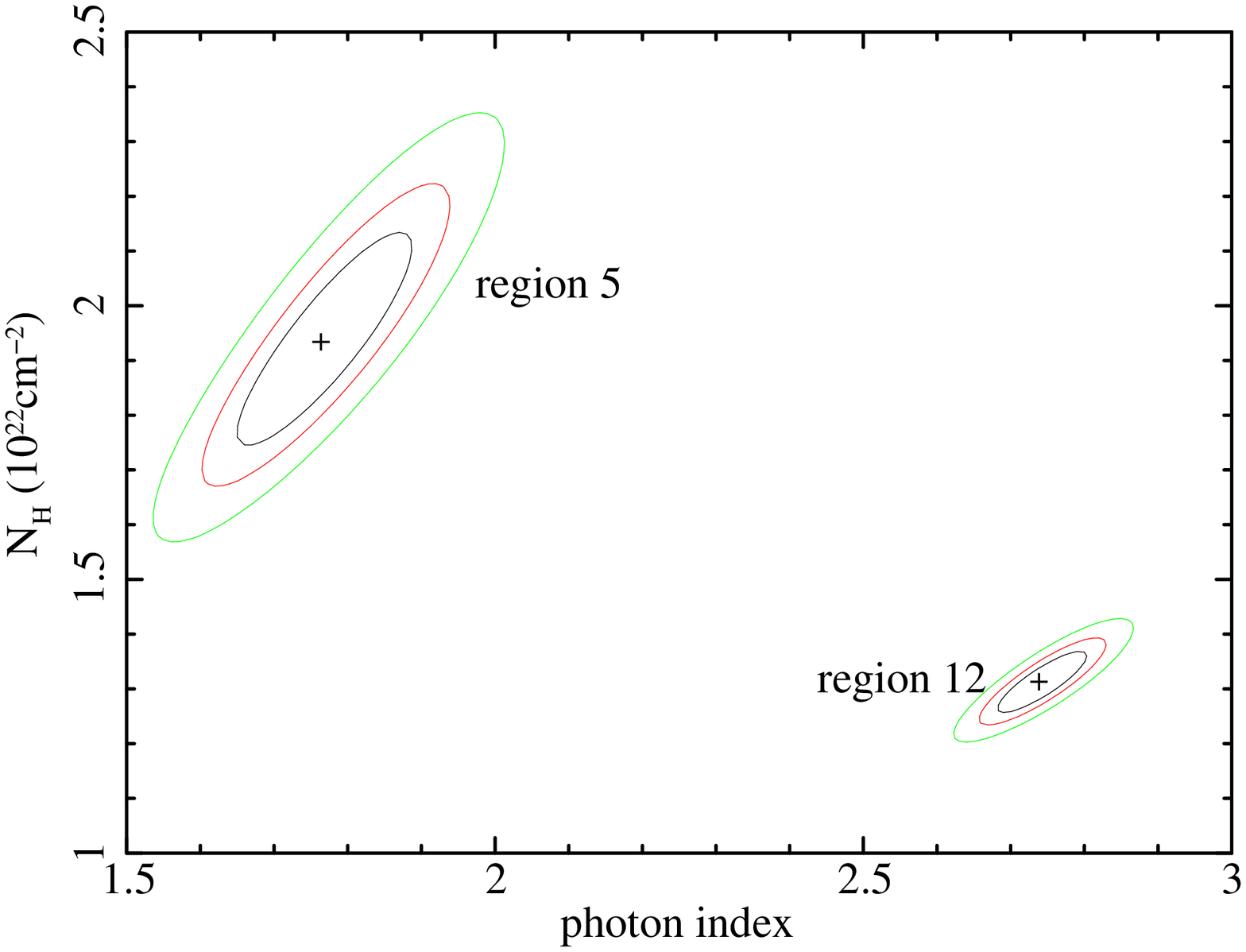}
\caption{
68\% (black), 90\% (red), and 99\% (green)
confidence contours of the photon index
vs. absorption column
for North-West, Center-East, and South regions (left)
and regions 5 and 12 (right).
}
\label{fig:contours}
\end{figure}

In a power-law spectrum,
photon index and absorption column densities are correlated to some extent.
To further verify that the change of $\Gamma$ and $N_{\rm H}$ in the 
{\it Suzaku} data across the remnant is indeed significant,
we compared the confidence contours of $\Gamma$ and $N_{\rm H}$
in the three cardinal areas.
The left panel of Fig.~\ref{fig:contours} confirms that
the North-Western area has significantly harder emission
and higher absorption
than the other two areas,
and the south area has the softest emission.
Similar checks were also done for several individual regions;
the right panel of Fig.~\ref{fig:contours}
illustrates that region 5 indeed exhibits significantly harder
and more absorbed emission than region 12.

To investigate whether the source emits thermal radiation,
we searched for emission lines of thermally heated plasma
on top of the continuum spectra, but could not identify promising candidates.
This applies to the spectra derived from individual regions,
as well as to the combined spectra from the three cardinal areas
discussed above.
We also tested the possibility that the lineless spectra
could come from a highly absorbed hot plasma with low ionization age
instead of from a nonthermal electron plasma.
For this purpose,
we fitted the softest spectrum from region 9
which also has the highest absorption
column density,
with an absorbed non-equilibrium ionization plasma model
\citep[NEI model,][]{borkowski2001}.
Indeed, the fit is statistically acceptable with a reduced $\chi^2$ of 47.2/39.
The resulting temperature is however higher than $\sim$2~keV,
with an ionization timescale of $\sim 10^9$~s~cm$^{-3}$.
Same conclusion or result
holds if several regions with soft spectra are combined,
using e.g. regions with photon indices larger than 2.5
in the sourthern half,
namely regions 7, 8, 9, 11, 12, and 13.
Again, the NEI model reproduces the spectra well 
with a reduced $\chi^2$ of 60.7/118,
with similar parameters as in region 9
($kT > $2.4~keV, 
ionization time scale less than $\sim 2\times 10^8$~s~cm$^{-3}$).
Such parameters would require a very young SNR or a recent encounter
of a fast shock with a dense region,
e.g. in a wind bubble scenario.
Since age and environment of HESS~J1731$-$347 are not known,
such an interpretation is formally not precluded.
It is however more natural to assume that
the spectra in these regions are dominated by nonthermal synchrotron emission,
in agreement with the hard regions where a pure thermal interpretation is
virtually excluded.

\subsubsection{Spectrum of XMMU~J173203.3$-$344518}

The central source of the SNR, XMMU~J173203.3$-$344518,
is in the field of view of the Suzaku XISs
only in the first Suzaku pointing (ObsID 401099010, 2007/02/23$-$24).
A first analysis of that data from the object has already been published
by \citet{tian2010}.
The uncertainty of their analysis is from
the combination of the relatively coarse angular resolution ($\sim$2~arcmin)
of Suzaku and
the contamination from the surrounding SNR emission.
We therefore repeated the spectral extraction,
adding systematic checks to verify the robustness of
the spectral results.

We used a circular source extraction region of 2.5~arcmin radius
centered on XMMU~J173203.3$-$344518
for the spectral analysis,
which encompasses $\sim$90\% of all source photons \citep{serlemitsos2007}.
As nominal background region,
we chose an annulus around the source
with 3--4.5~arcmin radius.

As already known,
the central source has a rather soft spectrum 
compared with the diffuse emission from the SNR.
As the first assumption, 
an absorbed blackbody model and an absorbed power-law model
were fit to the data.
The fits are statistically unacceptable
(reduced $\chi^2$ of 388.8/304 or 443.8/304, respectively).
We conclude that the spectrum has a multi component emission.

\begin{figure}
\epsscale{0.45}
\plotone{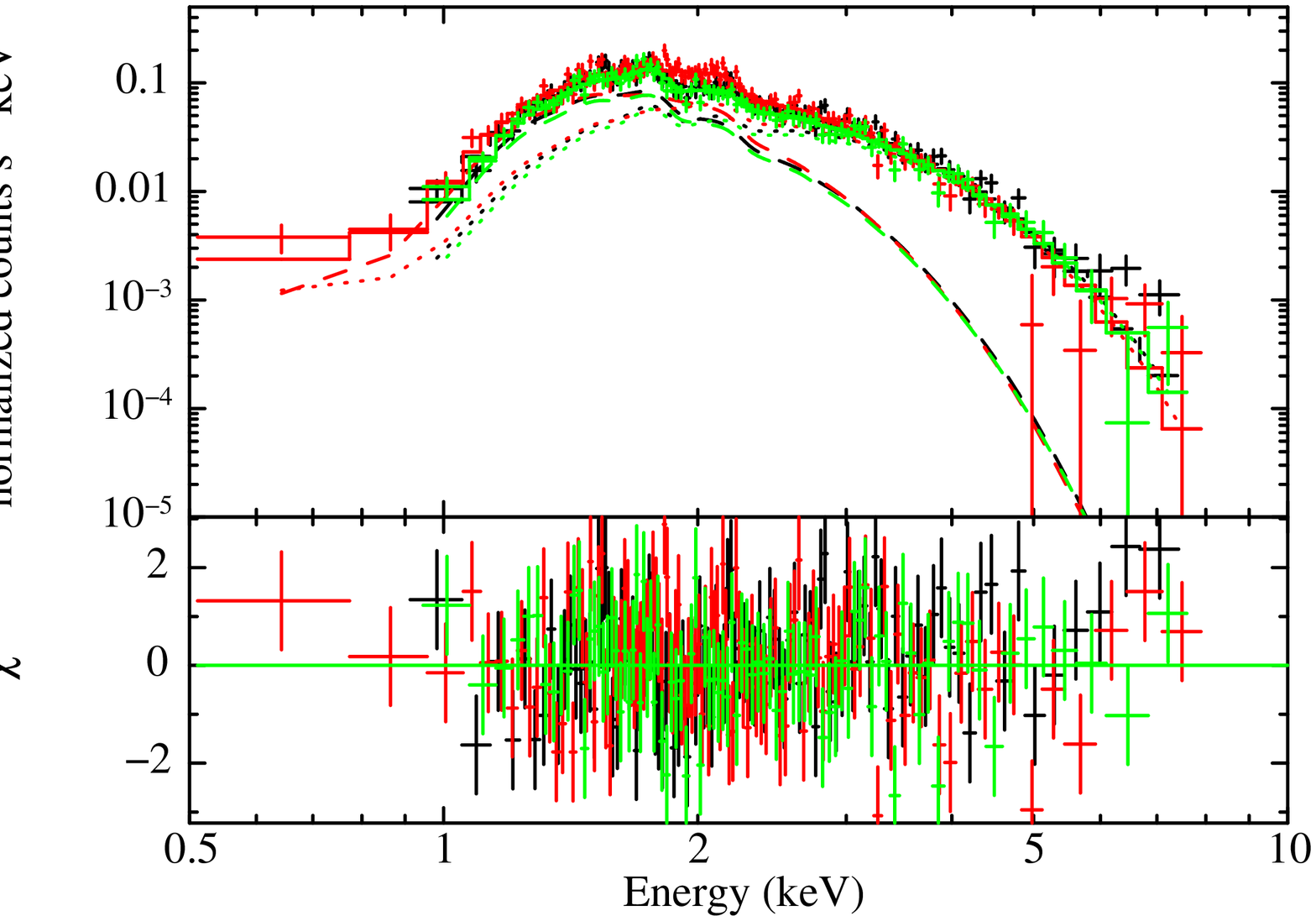}
\plotone{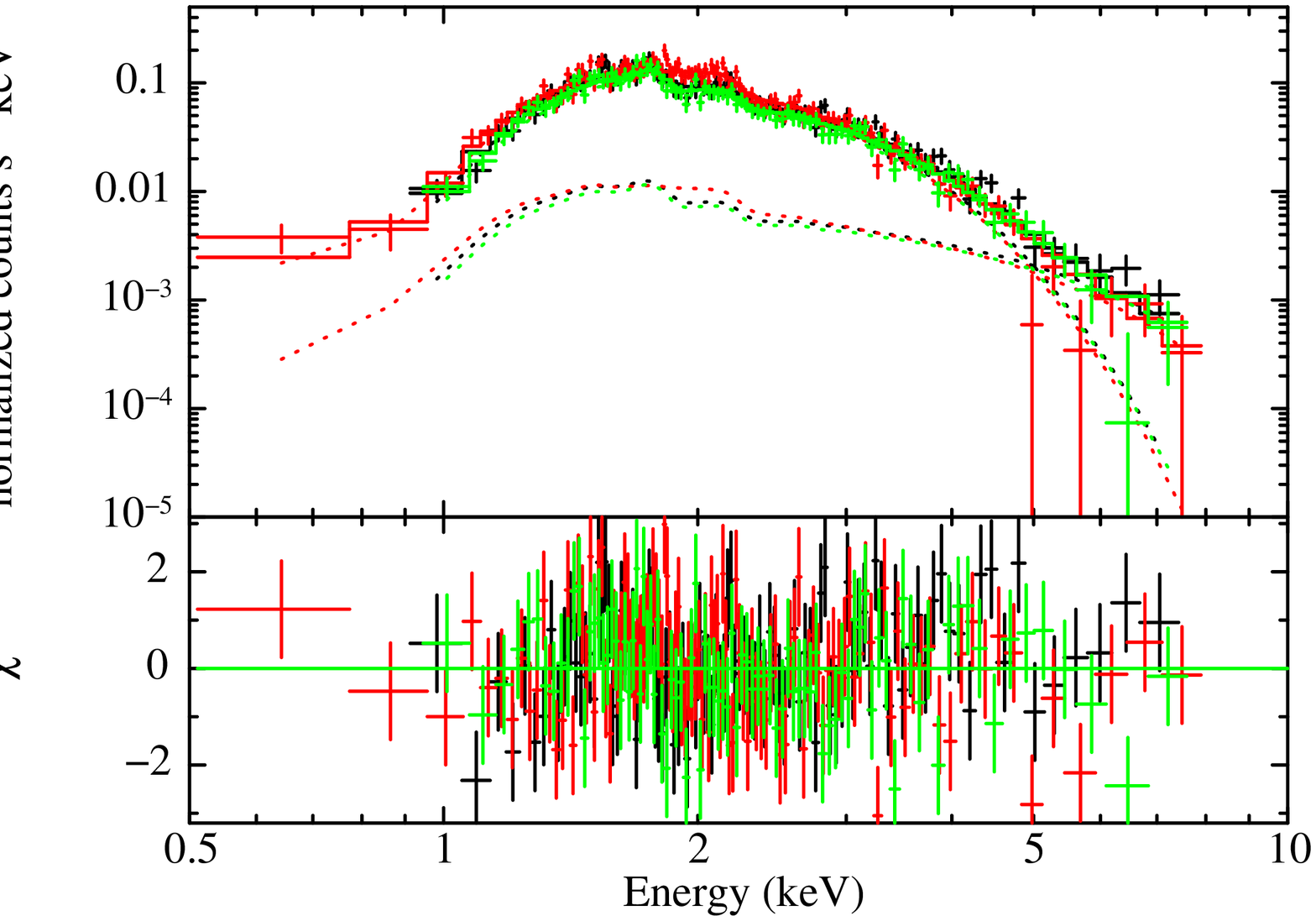}
\caption{XIS spectra of XMMU~J173203.3$-$344518.
The color represents XIS0 (black), XIS1 (red), and XIS3 (green).
Left panel shows the two-temperature model fitting,
whereas the right one shows a single blackbody plus 
SNR contamination model fitting.
The lower panels show the residuals from the best-fit model.
}
\label{fig:ccospectra}
\end{figure}

Fitting with
a blackbody plus a power-law component model
or a two-temperature blackbody model, yields better results,
with reduced $\chi^2$ of 334.7/302 or 340.7/302.
The best-fit parameters for the blackbody plus power-law model
are consistent with the results by \citet{tian2010}.
The photon index for the power-law component, $4.7_{-0.5}^{+0.5}$,
is rather soft for magnetars and CCOs;
we therefore used a two temperature blackbody model in this paper.
We show the spectrum with the best-fit model in Fig.~\ref{fig:ccospectra}
and the best-fit parameters in Tab.~\ref{tab:ccospectra}.

The second component, on the other hand,
could be the contamination of the SNR emission.
To check this possibility,
we also fitted the spectrum with
absorbed power-law or blackbody model
plus the spectrum of the region 3 as a contamination,
which is in the vicinity of the central source.
The normalization of the contamination component was
treated to be a free parameter,
whereas photon index and absorption column density are fixed
to those for region 3 spectrum.
The $\chi^2$/d.o.f. becomes 443.8/303 for the power-law case
and 350.8/303 for the blackbody case.
The latter case has similar $\chi^2$/d.o.f. as the multi-component models.
It is found that 
the source spectrum can be well reproduced by a single blackbody model
and the second component in the spectra could be
just the contamination of the diffuse SNR emission.
The best-fit parameters of the previous model
are shown in Tab.~\ref{tab:ccospectra}.



\subsection{Spectra above 10~keV}

We also analysed the HXD PIN data-set
to search for a signal from the source in the energy range above 10~keV.
After background (NXB+CXB) subtraction,
the remaining count rate in the 15--40~keV band
in each observation is 
$3.7\pm0.3\times 10^{-2}$ cnts~s$^{-1}$ for OBSID=401099010,
$1.7\pm0.3\times 10^{-2}$ cnts~s$^{-1}$ for OBSID=504031010,
and
$2.3\pm0.3\times 10^{-2}$ cnts~s$^{-1}$ for OBSID=504032010.
This is 12.4\%, 2.9\%, and 7.9\% of the NXB count rate
in this band,
respectively.
The systematic NXB uncertainty is 3--5\%
\citep{fukazawa2009},
i.e. $\sim$1--1.5$\times 10^{-2}$~cnts~s$^{-1}$,
but the uncertainty could be larger when the observation duration is
shorter than 1~day \citep{fukazawa2009}
as is the case here.
Ignoring for the moment this caveat,
the first and third observation appear to display
significant emission above instrumental and extragalactic backgrounds
in the 15--40~keV background.

Before potentially being able to attribute the excess signal
to HESS~J1731$-$347,
another background, 
the GRXE,
has to be considered.
Its level is difficult to estimate since it is rather faint
above 10~keV
and  its measurement is contaminated by bright point sources.
\citet{revnivtsev2003} revealed with {\it RXTE} that
the GRXE in this region is a few \% of the diffuse emission
in the Galactic center (GC) region.
This result is consistent with results obtained with {\it  GINGA}
X-ray satellite
(Yamauchi and Ebisawa, private communication).
With HXD PIN,
the GC diffuse emission was resolved 
to be $4\times 10^{-10}$~erg~cm$^{-2}$s$^{-1}$deg$^{-2}$
in the 12--40~keV band,
or $\sim 0.5$~cnt~s$^{-1}$ in the 10--40~keV band
\citep{yuasa2008}.
Our count rate is roughly 10\% of that in the GC region,
thus the emission is marginally above the GRXE.

Since HXD has no spatial resolution,
any point source in the field of view ($\sim$0.5~deg. FWHM for PIN)
adds to the total count rate in the detector.
First of all, the CCO, XMMU~J173203.3$-$344518,
is almost at the center of OBSID=401099010.
However, its spectrum is very soft,
well described by a thermal blackbody with temperature 0.5~keV
\citep{acero2009,tian2010,abramowski2011},
and does therefore not significantly contribute to the
emission above 10~keV
in the field.

To identify potential hard-band emitting sources,
we compared the three {\it Suzaku} fields of view
with the {\it INTEGRAL} Fourth IBIS/ISGRI Soft Gamma-ray Survey Catalog
\citep{bird2010}.
As a result, we found 4U~1728$-$337, a low-mass X-ray binary,
which is $\sim$50~arcmin away from the center of the SNR.
The source is variable with a 40--70~keV flux of
$\sim$(1--6)$\times 10^{-10}$~ergs~s$^{-1}$cm$^{-2}$
\citep{claret1994}.
During our {\it Suzaku} observations,
4U~1728$-$337 was detected with the RXTE All Sky Monitor (ASM)
with a count rate of a few cnts~s$^{-1}$.
Converting this value into an on-axis HXD PIN count rate results in
a rough count rate estimate of $\sim$0.1-1~cnts~s$^{-1}$
in the 15--40~keV band.
For the actual source location at the edge of the HXD PIN FOV
the expected count rate is one order of magnitude smaller,
i.e. $\sim$0.01-0.1~cnts~s$^{-1}$.
Given that this count rate is at the same level as the measured excess,
from our observations we cannot claim any detection of X-rays
above 10~keV from HESS~J1731$-$347.
Further studies using forthcoming {\it ASTRO-H} \citep{takahashi2010}
and/or {\it NuStar} \citep{harrison2010}
are needed to confirm this.

\subsection{Searching for coherent pulsations from the central compact object}

X-ray pulsations in the range of 0.1--0.4~s have been detected
from a few CCOs \citep{halpern2010c}.
On the other hand, magnetars typically exhibit pulsations
in the range of 2--12 seconds.
The fact that the claimed pulsation of $\sim$1~s \citep{halpern2010}
could not be confirmed \citep{halpern2010b} does not preclude
pulsations from XMMU~J173203.3$-$344518.

Below 10~keV, the time resolution of XIS is 8~s
\citep{koyama2007},
insufficient for a sensitive search for pulsations.
Above 10~keV, HXD provides adequate time resolution (61~$\mu$s),
and systematic background effects are negligible
when searching for coherent pulsations \citep{terada2008,terada2008b}.
Hence, even though the source was not significantly detected
in the HXD spectral analysis,
a pulsation search was still performed.
However, no significant pulsations were found in the scanned period range
of 0.1~s to 1000~s.

\section{Discussion}
\label{sec:discussion}

\subsection{Diffuse emission from SNRs}

The detailed analysis of the {\it Suzaku} data revealed that
the North-Western shell of HESS~J1731$-$347 exhibits stronger absorption
than the rest of the source.
The same trend has already been observed with {\it XMM-Newton}
\citep{abramowski2011}.
Actually, the statistics obtained with {\it Suzaku} permit to test
smaller scales,
but the two regions (4 and 9, see Fig.~\ref{fig:spectral_change})
that seem to deviate from the overall trend also have
the lowest surface brightness, and further investigation is needed
to verify the result.

Towards the North-West of the SNR lies the Galactic Plane.
However, the increase of the X-ray absorption column density can
most likely not be attributed to the average gas density increase
towards the Galactic Plane,
since the measured $N_{\rm H}$ (1.9--2.1$\times 10^{22}$~cm$^{-2}$)
is in clear excess of the average expected absorption column density
in this direction
(1.2--1.3$\times 10^{22}$~cm$^{-2}$ \citep{dickey1990,kalberla2005}).
That value is in fact consistent with the overall column density
measured towards the South-East of the source.
Indeed, \citet{abramowski2011} showed that
the absorption column derived from the {\it XMM-Newton} Eastern coverage
of the source is well correlated with molecular gas,
the density of which is derived from CO observations towards that region.
The question remains whether the SNR is located at a distance of $\sim$3~kpc
and interacting with a dense molecular cloud
identified through the CO measurement at that distance,
or whether the source is in fact located behind the cloud.
CO observations of higher excited states may be helpful
to find shock-heated gas. 

The non-uniformity of the
measured X-ray photon indices might offer another clue to that question.
In SNRs, hard synchrotron spectra such as the ones detected here
are typically attributed to electrons acceleration
in fast shocks (several thousand km~s$^{-1}$).
The harder the spectrum, the higher the maximum particle energy
and hence the higher the speed of the shock.
A lower shock speed might indicate the presence of dense material
into which the shock is propagating.
However,
in the observations presented here,
the North-West, which displays the highest column densities,
also shows the hardest photon indices
(e.g., region 1 and 5 in Table~\ref{tab:spectra}),
at odds with a cloud-interaction scenario.
We therefore stress again that at present there is no observational implication
of the SNR-molecular cloud interaction in the North-Western part,
and that it is more likely that
the absorbing material lies in front of the source.
Furthermore, the flat azimuthal VHE profile and
the lack of correlation between the absorption column and X-ray
surface brightness also indicate that the SNR may in fact not be
interacting with the molecular cloud but rather that the source is
located behind the cloud.

The accumulated unabsorbed X-ray flux in our analysed regions is
$2.3\times 10^{-11}$~ergs~cm$^{-2}$s$^{-1}$ in the 2--10~keV band.
For the entire remnant,
using our {\it Suzaku} observations together with the
{\it XMM-Newton} results \citep{abramowski2011},
the total flux of the source in this band is
approximately $\sim 5\times 10^{-11}$~erg~cm$^{-2}$s$^{-1}$.
The non-thermal X-ray luminosity $L$ and the radius of the source $R$
are estimated to
$L \sim 5\times 10^{34}$~erg~s$^{-1} \times d_{\rm 3kpc}{}^2$
and $R\sim 14~{\rm pc} \times d_{\rm 3kpc}$
($d_{\rm 3kpc}$ = $d$/(3~kpc), where $d$ is the distance to
the remnant).
For a distance of 3~kpc, these values are in rough agreement with
the SNR radius vs. non-thermal X-ray luminosity relation
of \citet{nakamura2011b}.
If this relation also holds for HESS~J1731$-$347,
then it is possible that the SNR is in fact
very near (but still behind) the molecular cloud 
that is responsible for the strong X-ray absorption.

Typical photon indices in the soft X-ray band from young SNRs
like RX~J1713$-$3946
range between $\sim$2.3 and $\sim$2.6 \citep{koyama1997,slane2001}.
In the North-West of HESS~J1731$-$347, $\Gamma$ is as small as
$\sim$1.8.
The only SNR known so far which shows a comparable photon index
is CTB~37B \citep{nakamura2009},
though the spectral extraction in the region of interest was 
very complex and subject to systematic uncertainties.
Recently, \citet{takahashi2008} discovered that
the synchrotron spectra of RX~J1713$-$3946 have a spectral break
in the hard X-ray band,
and that the photon index below the break is quite hard ($\sim$1.5).
A similar break was also found in SN~1006 \citep{bamba2008}.
We speculate that
the hard spectra in the North-Western region of HESS~J1731$-$347
with $\Gamma \le 2$ are of similar nature.  The Western and Southern
regions have softer X-ray spectra. The photon index of 2.3--3.0 is typical for 
synchrotron X-rays from SNR shells
\citep[][for example]{bamba2005a}.
We note that the Western part of the X-ray shell has a smaller curvature
than the North-Western and Southern shell segments.
This area looks promising to evaluate whether the VHE emission here
does follow this trend,
or whether the VHE shell is compatible with the larger average  shell radius
as the images seem to indicate (see Fig.~\ref{fig:multiwavelength}).
Such differences could be indicative of shock-cloud interaction
and might provide important clues of shock-accelerated protons in SNRs:
The X-rays trace the shock which has slowed down
due to the encounter with a molecular cloud,
while high energy protons have partially escaped the shock
and are primarily emitting VHE $\gamma$-rays
in the dense molecular cloud material.
Similar phenomena are seen in W28 north-eastern shell
\citep{nakamura2011} and G359.1$-$0.5 \citep{bamba2009}.
As discussed earlier in this paper,
the available VHE statistics are presumably not sufficient
for such a test at this point of time.
Also, the coverage of the X-ray emission in this particular area
should be extended
towards the West to pin-point the outermost X-ray boundary here.
%

In several SNRs, 
synchrotron X-ray spectra are softer in downstream regions
than regions close to the shock,
e.g. in SN~1006 \citep{rothenflug2004}
or in Tycho's SNR \citep{cassam-chenai2007}.
Within the still limited {\it Suzaku} coverage of HESS~J1731$-$347,
however, no systematic trend could be found 
in the photon indices when comparing regions
which are apparently shock-dominated with regions closer to
the center of the remnant.


\subsection{XMMU~J173203.3$-$344518}

The spectrum of the central source XMMU~J173203.3$-$344518
has at least one temperature blackbody component.
If the hard component is represented by a power-law,
the best-fit photon index,
4.7 (4.2--5.2), 
is not compatible with the hard tails usually detected in magnetars
\citep[c.f.,][]{enoto2010}.
Moreover,
the best-fit absorption column of 2.5 (2.2--2.7)$\times 10^{22}~{\rm cm}^2$
significantly exceeds the values measured from
the diffuse emission of the SNR.
We therefore
adopt in the following the two temperature blackbody model 
for the emission from this source.

Assuming that the distance to this source is 3~kpc,
the radius of the emitting region is 
2.2 (1.4--4.4)~km for the low $kT$ component ($\equiv R_{\rm LT}$)
and 0.6 (0.2--0.8)~km for the high $kT$ component
($\equiv R_{\rm HT}$), respectively,
i.e. the radiation is emitted from localized spots on the neutron star surface.

Although no flare has been detected from XMMU~J173203.3$-$344518,
it is instructive to compare the spectral results to flare spectra
of soft gamma-ray repeaters (SGRs).
These are well reproduced by two temperature blackbody models,
and the radii of the two components has a positive correlation of 
$R_{\rm HT}{}^2/R_{\rm LT}{}^2 \sim$0.03--0.4
\citep{nakagawa2007}.
For XMMU~J173203.3$-$344518,
the ratio is 0.06,
consistent with the above relation.
Therefore, although the best-fit temperatures of both components
are lower than flares,
it is suggestive to interpret
the emission from XMMU~J173203.3$-$344518
in a magnetar scenario.
Nevertheless, while a magnetar nature of XMMU~J173203.3$-$344518 is possible,
it clearly cannot be excluded,
that the source is a CCO associated with the supernova remnant.
Such objects are typically found in young SNRs and
characterized by soft black body-like emission with $kT\simeq 0.5$~keV
and a lack of variability 
(except pulsations with periods below one second detected in some cases).
The spectrum of one such object,
PSR~J1852+0040 --- a confirmed CCO in SNR Kes 79, has also been
fitted with a two-component black body ($kT=$0.3 and 0.52~keV),
as reported by \citet{halpern2010}.

On the other hand,
the harder component can be just contamination of the diffuse SNR emission.
The source spectrum in this case,
$\sim$0.5~keV blackbody, is also typical for CCOs.
Further detailed observations with good spatial resolution and statistics
are needed to confirm the nature of this source.

\acknowledgments

We thank the anonymous referee for the useful comments.
We thank all members of the {\it Suzaku} team
who kindly scheduled these observations.
We also thank S. Yamauchi and K. Ebisawa
for the helpful comments on the GXRE,
and Y. Fukazawa, T. Mizuno, and Y. Terada for the help of HXD analysis,
and S. Terada for the help of writing this paper.
This work was supported in part by
Grant-in-Aid for Scientific Research
of the Japanese Ministry of Education, Culture, Sports, Science
and Technology (MEXT) of Japan, No.~22684012 (A.~B.)
and No.~21740184 (R.~Y.).
W.W.T. acknowledges
support from NSFC, BeiRen Program of the CAS, and the 973 Program
(2012CB821800).

\begin{deluxetable}{p{5pc}ccccc}
\tabletypesize{\scriptsize}
\tablecaption{Observation Log
\label{tab:obslog}}
\tablewidth{0pt}
\tablehead{
\colhead{ObsID} & \colhead{Date} & \colhead{Position} & \colhead{XIS Exposure} & \colhead{HXD Exposure} & \colhead{SCI} \\
 & YYYY/MM/DD & (J2000) & [ksec] & [ksec]
}
\startdata
401099010\dotfill & 2007/02/23-24 & (263.0179, $-$34.7706) & 41 & 35 & OFF \\
504031010\dotfill & 2010/02/18-19 & (263.0074, $-$34.9458) & 42 & 34 & ON \\
504032010\dotfill & 2010/02/17-18 & (262.8483, $-$34.6325) & 42 & 33 & ON
\enddata
\end{deluxetable}

\begin{deluxetable}{p{4pc}ccccc}
\tabletypesize{\scriptsize}
\tablecaption{Best-fit parameters of spectral fittings\tablenotemark{a}.
\label{tab:spectra}}
\tablewidth{0pt}
\tablehead{
\colhead{Region} &
\colhead{$N_{\rm H}$} &
\colhead{$\Gamma$} &
\colhead{$F_{\rm 2-10 keV}$} &
\colhead{Surface brightness} &
\colhead{$\chi^2$/d.o.f.}\\
& ($10^{22}$~cm$^{-2}$) & & ($10^{-12}$~erg~cm$^{-2}$s$^{-1}$) & 
($10^{-13}$~erg~cm$^{-2}$s$^{-1}$arcmin$^{-2}$)
}
\startdata
1\dotfill & 2.11 (1.92--2.30) & 2.21 (2.09--2.33) & 2.07 (2.00--2.15) & 1.04 (1.00--1.08) & 173.6/153 \\
2\dotfill & 1.78 (1.61--1.95) & 2.68 (2.53--2.84) & 1.66 (1.61--1.95) & 1.38 (1.34--1.63) & 98.5/93 \\
3\dotfill & 1.40 (1.34--1.47) & 2.37 (2.31--2.43) & 5.44 (5.32--5.56) & 2.72 (2.66--2.78) & 371.5/320 \\
4\dotfill & 1.12 (0.92--1.36) & 2.30 (2.09--2.54) & 0.73 (0.67--0.78) & 0.42 (0.38--0.45) & 111.7/79 \\
5\dotfill & 1.93 (1.73--2.15) & 1.77 (1.64--1.90) & 2.19 (2.10--2.27) & 1.46 (1.40--1.51) & 144.2/124 \\
6\dotfill & 1.11 (1.02--1.21) & 2.42 (2.32--2.53) & 2.01 (1.94--2.09) & 1.68 (1.62--1.74) & 137.8/136 \\
7\dotfill & 0.96 (0.82--1.12) & 2.59 (2.38--2.81) & 1.17 (1.05--1.29) & 1.11 (1.00--1.23) & 54.8/60 \\
8\dotfill & 1.15 (1.00--1.31) & 2.52 (2.33--2.71) & 1.54 (1.43--1.66) & 0.88 (0.82--0.95) & 119.0/94 \\
9\dotfill & 2.52 (1.96--3.18) & 3.17 (2.70--3.71) & 0.19 (0.13--0.24) & 0.16 (0.11--0.20) & 47.8/40 \\
10\dotfill & 1.63 (1.42--1.86) & 2.48 (2.29--2.68) & 1.56 (1.43--1.69) & 1.27 (1.17--1.38) & 49.7/74 \\
11\dotfill & 1.10 (1.00--1.20) & 2.54 (2.43--2.66) & 1.53 (1.47--1.60) & 1.25 (1.20--1.31) & 122.7/124 \\
12\dotfill & 1.31 (1.25--1.37) & 2.74 (2.68--2.81) & 3.20 (3.12--3.27) & 1.31 (1.27--1.33) & 380.4/337 \\
13\dotfill & 1.40 (1.18--1.65) & 2.55 (2.34--2.78) & 0.65 (0.60--0.70) & 0.53 (0.49--0.57) & 82.5/65 \\
North-West\dotfill & 1.66 (1.54--1.79) & 1.97 (1.88--2.06) & 4.55 (4.43--4.67) & 0.86 (0.84--0.89) & 240.2/180 \\
Center-East\dotfill & 1.29 (1.24--1.35) & 2.36 (2.30--2.41) & 7.53 (7.38--7.68) & 2.35 (2.31--2.40) & 318.0/229 \\
South\dotfill & 1.30 (1.24--1.36) & 2.63 (2.56--2.70) & 7.74 (7.50--7.98) & 0.85 (0.82--0.88) & 347.4/355
\enddata
\tablenotetext{a}{Errors indicate single parameter 90\% confidence regions.}
\end{deluxetable}

\begin{deluxetable}{p{20pc}cc}
\tabletypesize{\scriptsize}
\tablecaption{Best-fit parameters of spectral fittings
of XMMU~J173203.3$-$344518\tablenotemark{a}.
\label{tab:ccospectra}}
\tablewidth{0pt}
\tablehead{
\colhead{Parameters} & \colhead{2 blackbody case} & \colhead{blackbody + SNR contamination}
}
\startdata
$N_{\rm H}$ ($10^{22}$~cm$^{-2}$)\dotfill & 1.7 (1.5--2.0) & 1.5 (1.4--1.6) \\
$kT_1$ (keV)\dotfill & 0.33 (0.26--0.42) & 0.47 (0.46--0.48)\\
$Norm_1$\tablenotemark{b}\dotfill & 14 (12--16) \\
$kT_2$ (keV)\dotfill & 0.58 (0.53--0.72) & --- \\ 
$Norm_2$\tablenotemark{b}\dotfill & 3.5 (0.6--7.0) & --- \\
SNR contamination ($10^{-13}$~erg~cm$^{-2}$s$^{-1}$arcmin$^{-2}$)\tablenotemark{c}\dotfill & --- & 5.6 (4.2--7.0)
\enddata
\tablenotetext{a}{Errors indicate single parameter 90\% confidence regions.}
\tablenotetext{b}{$R_{\rm km}{}^2/D_{10}{}^2$,
where $R_{\rm km}$ is the source radius in km,
and, $D_{10}$ is the distance to the source in units of 10~kpc.}
\tablenotetext{c}{2--10~keV band.}
\end{deluxetable}

\end{document}